\documentclass[twocolumn,floatfix]{aastex63}

\usepackage{color}

\received{\today}

\begin{document}

\shortauthors{Bronzwaer \& Falcke}

\title{THE NATURE OF BLACK HOLE SHADOWS}

\correspondingauthor{Thomas Bronzwaer}
\email{t.bronzwaer@astro.ru.nl}

\author{Thomas Bronzwaer}
\affiliation{Department of Astrophysics/IMAPP, Radboud University Nijmegen \\
                  P.O. Box 9010, 6500 GL Nijmegen, The Netherlands}
\author{Heino Falcke}
\affiliation{Department of Astrophysics/IMAPP, Radboud University Nijmegen \\
                  P.O. Box 9010, 6500 GL Nijmegen, The Netherlands}

\begin{abstract}
A distinct visual signature occurs in black holes that are surrounded by optically thin and geometrically thick emission regions. This signature is a sharp-edged dip in brightness that is coincident with the black-hole shadow, which is the projection of the black hole's unstable-photon region on the observer's sky. We highlight two key mechanisms responsible for producing the sharp-edged dip: i) the reduction of intensity observed in rays that intersect the unstable-photon region, and thus the perfectly absorbing event horizon, versus rays that do not (blocking), and ii) the increase of intensity observed in rays that travel along extended, horizon-circling paths near the boundary of the unstable-photon region (path-lengthening). 
We demonstrate that the black-hole shadow is a distinct phenomenon from the photon ring, and that models exist in which the former may be observed, but not the latter. Additionally, we show that the black-hole shadow and its associated visual signature differ from the more model-dependent brightness depressions associated with thin-disk models, because for geometrically thick and optically thin emission regions, the blocking and path-lengthening effects are quite general. Consequentially, the black-hole shadow is a robust and fairly model-independent observable for accreting black holes that are in the deep sub-Eddington regime, such as low-luminosity active galactic nuclei (LLAGN).

\end{abstract}


\section{Introduction} \label{sec:intro}



\subsection{The black-hole shadow}

In the recreational practice of coin rubbing, a sheet of paper is placed over a coin, and a pencil or crayon is rubbed on the paper. An impression of the coin appears in the rubbing. As the trick is repeated, the rubbings will look different each time, depending on such factors as the type and color of crayon, and the vigor of the rubbing; but, as long as a few basic rules are observed, the impression of the coin will always be the same. 

An analogy can be drawn with the appearance of a black hole, which itself provides a constant spacetime structure, but which can be illuminated by external sources of electromagnetic radiation, which may have a variety of shapes, emit a variety of colors, and behave in a time-dependent manner. Like the coin underneath a sheet of paper, the black hole itself cannot be seen, but its imprint on the light distribution \emph{can} be seen. Recently, the Event Horizon Telescope collaboration published the first image of a black hole \citep{eht1}, which exhibits this effect of a constant spacetime structure illuminated by a time-varying emission region, prompting us to better understand and explain what we see, and what we do not see.

The visual appearance of black holes was first studied by \citet{cunninghambardeen} and \citet{bardeen1974} for the case of a  star orbiting a black hole, as well as various related scenarios. This early work revealed the basic structure of a thin disk with a gravitationally lensed inner region. The first computer-calculated, but hand-draw, visualization of a black hole surrounded by a luminous accretion disk was presented by \citet{luminet1979}. These early models focused on geometrically thin, optically thick accretion disks (\citealt{shakura1973}; \citealt{novikovthorne1973}; \citealt{pagethorne}) that have so far dominated our perception of black holes in science and popular culture. Such accretion disks may be found in black-hole binary systems and quasi-stellar objects (QSO's) (see, e.g., \citet{agol1999z}). Those sources were found and extensively discussed in the 1960's and 1970's.  However, the vast majority of black holes in the universe  -- the silent majority \citep{falcke2001} -- operate at highly sub-Eddington accretion rates. These Low-Luminosity AGN (LLAGN) \citep{ho2008} are characterized by advection-dominated flows that are radiatively inefficient (see \citet{narayan1998} for a review), coupled to ubiquitous radio jets \citep{nagar2001}, both of which are expected to be geometrically thick and optically thin at some emission frequency. In contrast to Quasars and high-state X-ray binaries, which show a plethora of appearances and states, Sub-Eddington black holes are reasonably well-behaved \citep{FenderBelloniGallo2004a}, and the scaling of their spectral energy distribution is, to first order, described by a “fundamental plane relation” (\citealt{merloni2003}; \citealt{falcke2004}; \citealt{plotkin2012}) that represents the ratio between optically thick (typically radio) and optically thin (typically X-ray) emission as a function of black hole mass and accretion rate. The fundamental plane relation also determines the frequency at which the emission region surrounding a black hole becomes optically thin. This model explains why supermassive black holes are optically thin somewhere between the radio and NIR frequency regimes, while X-ray binaries typically become optically thin only in the NIR-X-ray regime. For a low-power supermassive black hole, like Sgr A* in the center of the Milky Way, optically thin emission occurs at mm-wavelengths, and approaches the scale of the event horizon \citep{FalckeMannheimBiermann1993a,NarayanYiMahadevan1995,FalckeGossMatsuo1998,OzelPsaltisNarayan2000}.

The difference between optically thin and optically thick emission regions is conceptually important, because optical-depth effects can significantly influence the appearance of emission regions around black holes \citep{bronzwaer2020}. The presence of a geometrically thick, optically thin emission region directly surrounding a black hole's event horizon gave rise to the concept of the black-hole shadow \citep{falcke2000} that is observed when imaging such accreting black holes, and whose features depend only on spacetime/observer geometry \citep{Johannsen2016b,JohannsenPsaltis2010a,bronzwaer2020}. The concept was derived from the work by \cite{Bardeen1973a}, who calculated the optical appearance of a black hole in front of a plane-parallel, uniformly radiating disk. While this scenario, e.g., a black hole in front of a star, is impossible to observe with current technology, the key visual element of this appearance is the so-called photon sphere - the boundary of the region around the black hole in which closed, circular photon orbits exist. It is this surface that determines the appearance of the black-hole shadow\footnote{Some of the terms that are used frequently by the Event Horizon Telescope (EHT), such as `photon ring’ and `black-hole shadow’, have been assigned multiple interpretations in the literature. In order to be consistent with our fellow EHT authors, we will henceforth adopt the terminology used by the EHT Gravitational Physics Inputs Working Group (GPIWG). In particular, we take `black hole shadow’ (BHS or shadow) to refer to the projection, on an observer’s sky, of the region around a black hole in which circular orbits exist; this region contains the set of observer viewing rays that cross the black hole's event horizon. The bounding curve of the shadow has also been called the `$n=\infty$ photon ring’ and the `critical curve’. Our definition of `photon ring’ is the system of lensed images of the black hole’s environment that, to the observer, appear just outside of the BHS. Finally, by `central brightness depression' (CBD) we refer to the characteristic central dip in brightness observed in black holes surrounded by optically thin accretion disks. As we shall show in this paper, systems exist in which the CBD does not match the BHS, but in which a distinct visual signature is nonetheless observable at the BHS.}.

Although a black hole's surrounding emission region, which renders the shadow visible in the first place, depends on plasma astrophysics and can have a highly time-dependent appearance (\citealt{falcke2001b, broderick2006a, noble2007, moscibrodzka2009}), the photon sphere - and thus the shadow itself - is as constant as the outline of a coin. As long as the emission region is optically thin and geometrically thick, a strong visual signature (a dip in intensity) that is coincident with the shadow should be observable. Moreover, the shadow's radius is proportional to the black hole's mass and, under favorable circumstances, the shadow's shape is informative about the black hole's spin (see, e.g., \citealt{johansen2010, younsi2016}). For these reasons, observing the black-hole shadow was a key driver for the Event Horizon Telescope (EHT, \citealt{EHT2019II}) and one of its clearest predictions \citep{falcke2000,broderick2006a,KamruddinDexter2013a,MoscibrodzkaFalckeShiokawa2016a}.

\subsection{Criticism of the black-hole shadow}

In the scrutiny of the EHT result subsequent to its publishing, criticism has been raised regarding the claim of having observed the black-hole shadow. The use of the term shadow is also occasionally questioned. \cite{Bardeen1973a} called his dark region the ``apparent shape of the black hole''.  \cite{Cash:2002tl} used the term silhouette, in the context of a proposed X-ray interferometer to observe accretion disks of X-ray binaries.  \citet{gralla2020}  used the term ``critical curve'' to describe the projection of the photon sphere's boundary on the observer's sky. The term shadow was introduced by \citet{falcke2000} and independently by \citet{devries2000} a few weeks later, and has been adopted by the Event Horizon Telescope (EHT) and a majority of the papers that have been published on the topic in the past two decades, both on the observational side and on the theoretical side (see, e.g., \citet{theoryref1,theoryref2,theoryref3}). 
While precise terminology is important, words derived from everyday human experience do not have a clear mathematical definition, nor do they have an equivalent in the realm of curved spacetimes. But they do carry a deeper meaning. Clearly, there is no real bang in the ``big bang'', and it is not so clear where the ``hole'' is in a black hole\footnote{For more on the etymology of the word ``black hole'', see \citet{blackholename}.}; still, the general public, as well as the scientific community, intuitively know what these terms mean. Hence, we here continue to use the term shadow for the geometric shape of the projection of the photon sphere on an observer's sky, a shape that corresponds to a characteristic CBD in low-power black holes.

Besides the semantics of our terminology, \citet{gralla2019} pointed out that certain emission models give rise to CBD's whose shapes are determined by the characteristics of the emission region, rather than the spacetime geometry; in some of these models, no strong visual signature is seen at the BHS. More generally, a concern exists that the EHT has produced images of the detailed astrophysical circumstances of the accretion disk/jet system, rather than of the black-hole shadow as an observable that is directly informative about the black hole's spacetime geometry. Additionally, it is claimed in \citet{gralla2020} that the black-hole shadow, as an observable feature, is ``not generic'', but that the \emph{photon ring} - a term which has been defined in several ways, but which we here take to mean: the complex system of lensed images of the emission region which nearly coincides with the boundary of the shadow - \emph{is} generic. Finally, it is claimed in the same paper that ``the black-hole shadow does not occur in the EHT models of M87*''. Taken together, these claims pull into doubt the validity the black-hole shadow, both as an astrophysical observable and as a theoretical concept. It is therefore important that they be addressed. 

\citet{gralla2019} presents several examples of emission regions whose CBD's deviate from the black-hole shadow. These may be roughly divided into two categories: the first category describes `backlit' black holes which are illuminated by distant screens, as in \citep{bardeen1974}, and the second category are optically and geometrically thin disks. The former class of models cannot reproduce the M87* image, while the latter category can, provided the disk is viewed in a nearly face-on manner. These models help us to understand why the geometrical thickness is a key ingredient for observing the shadow. This will be explored further in the coming sections. 

It is important to note that a wide range of reasonable astrophysical models has been investigated by the EHT Collaboration \citep{EHT2019VI}, as well as in \citep{bronzwaer2020} -- an endeavor that we do not seek to repeat here. These simulations include jets and disc structures of various sizes for a wider range of parameters, and they display a wide range of observed source morphologies. Some models, observed at a frequency at which the model is (partially) optically thick, show no visible shadow at all. Some show a partially obscured shadow, others an apparently exaggerated shadow. An interesting class of models appears to show a dip in brightness that is attributable to a direct image of the event horizon \citep{chael2021}.
In short, the overall appearance of the source is affected by a number of factors related to the emission region. The immutable shadow itself, however, can still be observed clearly for compact LLAGN sources, at frequencies at which the emission region is optically thin. 

In this work, we try to make more intuitively clear why, and under which conditions, the black-hole shadow is a robust and meaningful feature. We explore the physical processes that give rise to the observability of the shadow and show why it is that the shadow is in fact the more generic feature for geometrically extended emission regions. We also investigate the nature of the shadow in intermediate cases, such as toroidal and thin-disk emitters, which cover a range of astrophysical cases.

\subsection{Outline}

In Section \ref{sec:theory}, we analyze the visual appearance of dark objects surrounded by optically thin emission regions theoretically, first in a Newtonian approximation (with no light bending), and then in a general-relativistic context; we identify the key physical effects that cause us to observe a darkened region on the sky. Section \ref{sec:numerical} presents numerically computed image maps of black holes surrounded by various types of emitters, which demonstrate how the shadow behaves. We summarize our findings in Section \ref{sec:conclusions}.

\section{Radiative transfer: theory}
\label{sec:theory}

In this section, we outline a geometrical-optics treatment of the appearance of an opaque, absorbing, spherical object located within a luminous, optically thin cloud. We first examine the classical, Newtonian case, from which the effect of gravitational lensing is absent, and then move to the general-relativistic treatment. This allows us to better understand which effects produce the shadow, as well as which effects are specifically caused by the bending of light rays in GR.

\subsection{Newtonian case}
\begin{figure}[htp]
\centering
\epsscale{0.96}
\plotone{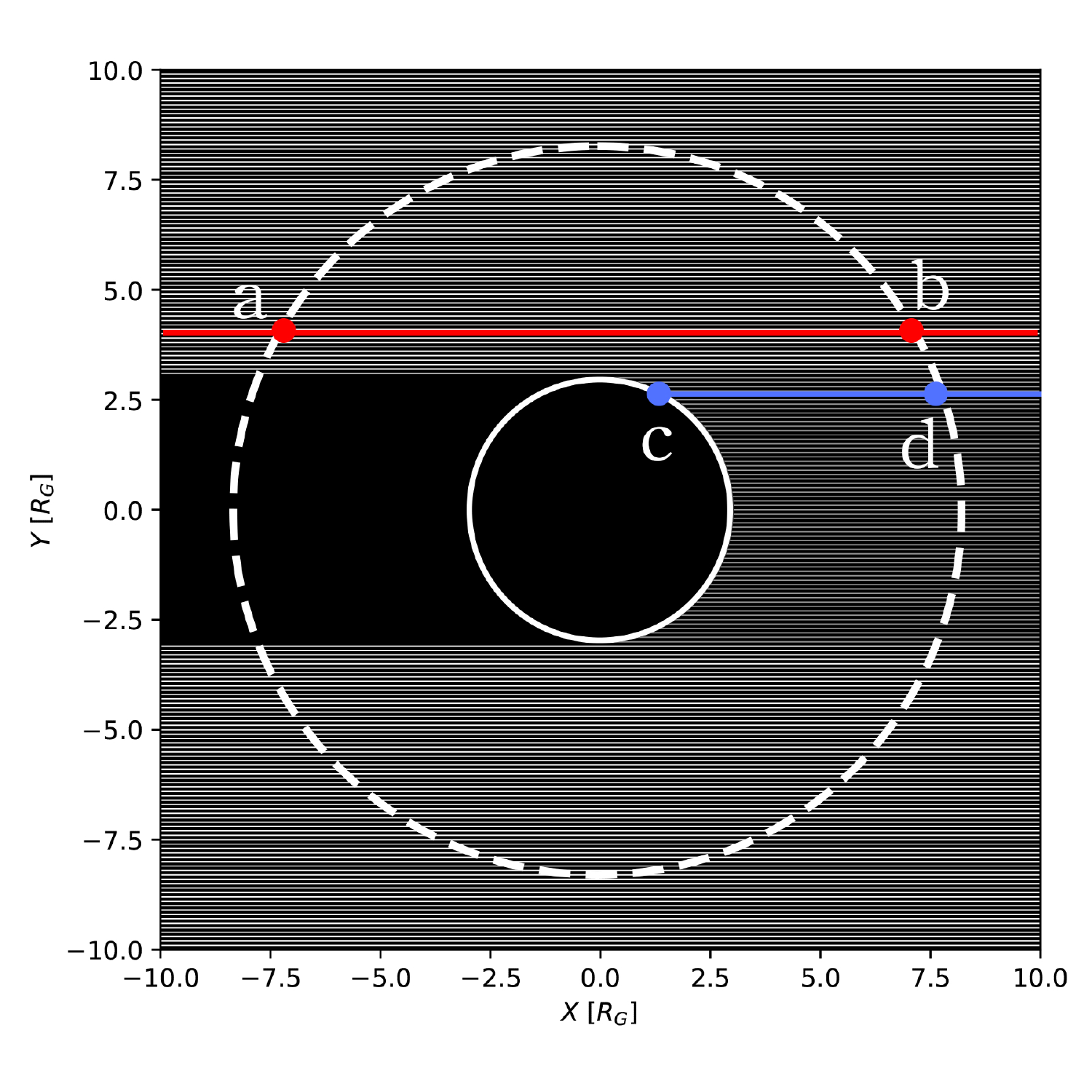}
\figcaption{An illustration of the blocking effect with Newtonian physics. No radiation emitted behind the opaque sphere (solid white circle) can reach the observer, so that rays that cross the opaque sphere (grey and blue lines) have reduced path lengths inside an emission region (dashed white circle) than rays that do not cross the opaque sphere (white and red lines).  \label{fig:blockedregion}}
\end{figure}

Consider a luminous cloud of roughly spherical shape, viewed by a distant observer, so that the viewing rays are practically parallel upon arrival.  A spherical, opaque object is added at the cloud's center. Given a particular observer's viewing direction, a region now exists from which no radiation can reach that observer (see Fig.~\ref{fig:blockedregion}). In everyday life such a situation, albeit with a more complex geometry, occurs if we look at the wick inside the flame of a burning candle or a kettle inside a raging campfire. It will be seen as a dark contour.

In Newtonian physics, the observer's viewing rays travel along straight lines, and an observer image is constructed by solving the radiative transfer equation those rays:
\begin{equation}
\frac{dI_{\nu}}{ds} = j_{\nu} - \alpha_{\nu} I_{\nu},
\label{eq:RTE}
\end{equation}
where $j$ is the emission coefficient (which is constant in our simple model), $I_{\nu}$ is the ray's specific intensity, and $\alpha_{\nu}$ is the absorption coefficient (which is zero in our model). By considering two rays, one of which passes through the sphere and the other doesn't (the blue and red lines in Fig.~\ref{fig:blockedregion}, respectively), we define the `darkening factor' $\chi$ as the ratio of the intensities of the two rays as they reach the observer: 
\begin{equation}
    \chi_{\nu} = \frac{\int_c^d j_{\nu} \ ds + I_{\nu,{\rm sph}}}{\int_a^b j_{\nu} \ ds},
    \label{eq:darkening_factor}
\end{equation}
where $s$ parametrizes the integration distance along the rays, $I_{\rm sph}$ is the specific intensity emitted by the sphere, and the position markers $a,b,c,d$ are indicated in Fig.~\ref{fig:blockedregion}. Note that $\chi_{\nu}$ is frequency-dependent. Therefore, we assume that the cloud is observed at a frequency $\nu$ at which it is optically thin.

In the case of a perfectly black sphere, $I_{\rm sph} \approx 0$ for all frequencies. Of course, this dark sphere occurs due to infinite redshift at the event horizon in GR, and thus would not occur in a Newtonian context. The point of this model is only to illustrate the principles that cause us to see a shadow. The sphere's location is near the cloud's center; this is a reasonable assumption, given the fact that the sphere, and not the optically thin cloud, will dominate the local gravitational field. In fact, the dominant central gravitational potential will force every reasonable emission region to either rotate or fall rapidly, which will smooth out plasma inhomogeneities over time and force the emission region to be symmetric around the central potential after a few dynamical timescales. Since we assume that the emission coefficient $j$ of our model is isotropic and constant along the rays, $\int_c^d j \ ds \approx 1/2 \int_a^b j \ ds$, so that we have
\begin{equation}
    \chi_{\rm bh} \approx \frac{1}{2}.
    \label{eq:chi_bh}
\end{equation}
Note that this is an upper bound on $\chi_{\rm bh}$ in realistic scenarios; for very small emitters, grazing rays will have much longer path lengths inside the emission region versus plunging rays, the arguments mentioned above will break, and the darkening will in fact be more severe, as the ratio of Eq.~\ref{eq:darkening_factor} diminishes in magnitude.

Crucially, in the case of an opaque sphere, the darkening factor will apply if and only if a ray intersects the sphere. The darkening factor thus appears as a step function in the image, in other words, the darkened region has a `sharp edge'. 

To sum up, in the Newtonian case, we should expect an optically thin cloud of arbitrary shape, with a black, opaque sphere near its center, to display a ``shadow'', a sharp-edged, darkened -- but not dark -- circle whose flux density is roughly half of the flux density from the surrounding regions. The reason we see this darkened circle is because the sphere blocks radiation from an approximately cylindrical region from reaching the observer. We shall call this the `blocking effect'.

We note that the above considerations hold for emitters of many different shapes (e.g., toroidal), as long as the sphere is completely occluded by the emitter; only the darkening factor is then affected by the instantaneous, detailed state of the emission region, while the outline of the black-hole shadow remains constant. Consider, for example, a toroidal emitter with a radius equal to the radius of the central sphere, viewed from within the equatorial plane. In this case the shadow will be visible, however, if we now move the observer, so that the torus is viewed in a face-on manner, the blocking effect vanishes and we do not see a shadow, but the hole in the torus. Given a significantly flattened emitter, the blocking effect is conditional. In the extreme case of a purely two-dimensional emitter, such as the Novikov-Thorne disk (\citealt{shakura1973}; \citealt{novikovthorne1973}), the blocking effect is also suppressed, and the outline of the sphere could at most be partially revealed in the Newtonian case (see Section \ref{sec:torus}).

\subsection{General-relativistic case}

How does the preceding discussion change if we move to a general-relativistic treatment? In Einstein's general theory of relativity (GR), the structure of spacetime is expressed using the metric metric tensor $g_{\mu \nu}$. Light rays move along null geodesics; in a powerful gravitational field (such as that near the center of the Kerr metric considered in this paper), null geodesics deviate from straight lines, and become curved. A light ray's direction of travel is characterized by the wave vector $k^{\mu}$, which is null-normalized: $k^{\alpha} k_{\alpha}=0$. By parallel-transporting $k^{\mu}$ along itself, we obtain the geodesic equation:
\begin{equation}
k^{\alpha} \nabla_{\alpha} k^{\mu} = 0,
\label{eq:GE}
\end{equation}
where $\nabla_{\mu}$ is the covariant derivative.

In order to create a virtual image of a black hole, \ref{eq:GE} is solved for initial (camera) rays, and then, given a radiative model for an emitting source (e.g., an accretion disk or jet), radiative-transfer calculations are performed along the rays. To do so, we employ Lorentz-invariant versions of the intensity $\mathcal{I} = I_{\nu}/\nu^3$, the invariant emission coefficient $\mathcal{J}=j_{\nu} / \nu^2$, and the invariant absorption coefficient $\mathcal{A}=\alpha_{\nu} \nu$. We may then solve the invariant radiative-transfer equation along a ray:
\begin{equation}
    \frac{d\mathcal{I}}{d\lambda} = \mathcal{J} - \mathcal{A} \mathcal{I},
\end{equation}
and simply transform back to the usual quantities after integration.

\begin{figure}[htp]
\centering
\plotone{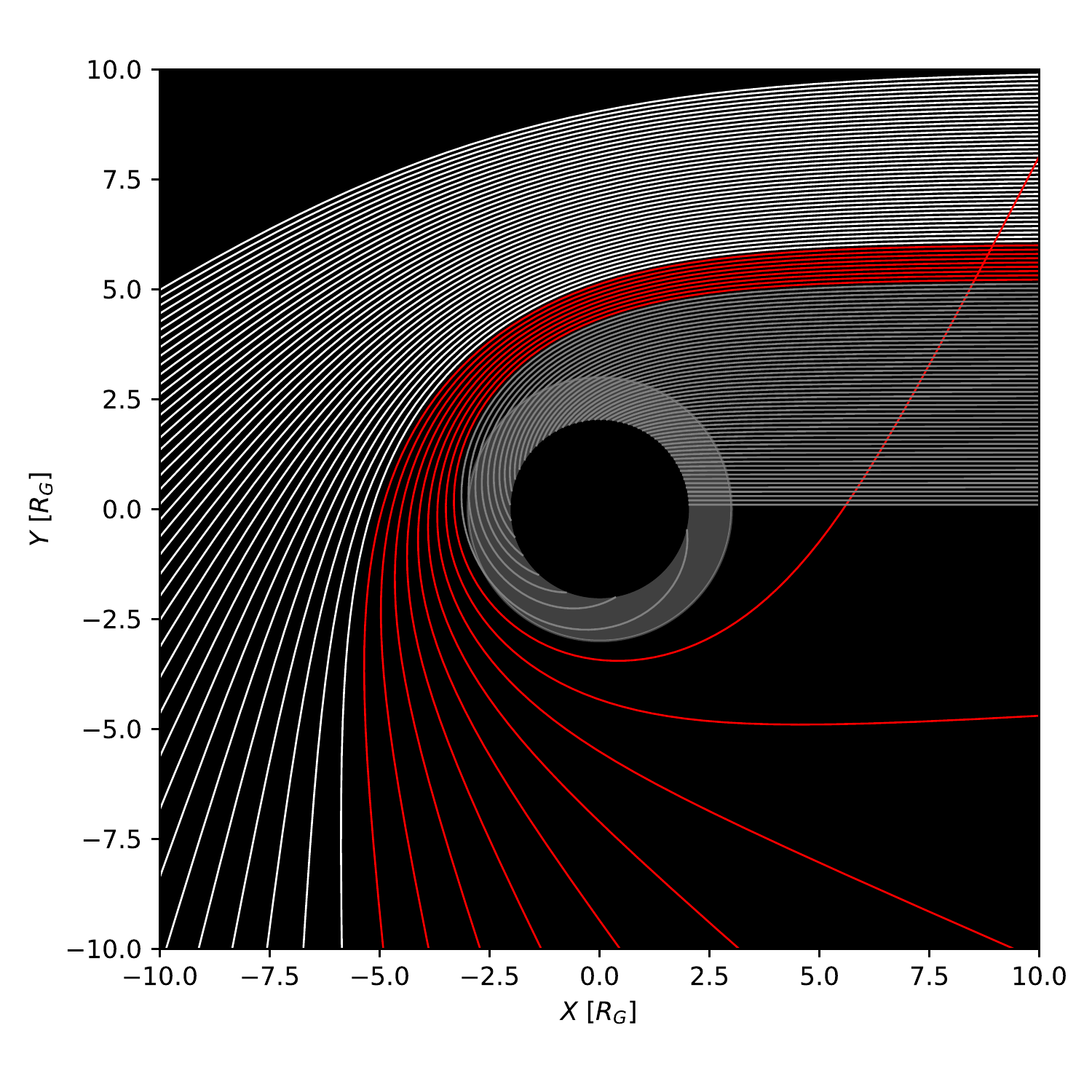}
\figcaption{A number of null geodesics in the equatorial plane of the Schwarzschild spacetime. The red and white geodesics do not cross the photon sphere, while the grey geodesics do, causing their path length to be reduced, and hence, the observer to see less flux in the corresponding viewing directions, as in the Newtonian case (Fig.~\ref{fig:blockedregion}). The red rays, which barely escape, are the set of rays with at least one turning point. Their paths are lengthened, causing the characteristic increase in brightness observed just outside the shadow. The solid grey circle marks the unstable-photon region; the solid black circle marks the event horizon.\label{fig:pathlengthening}}
\end{figure}

A key change that is introduced with respect to the Newtonian case is the gravitational lens (i.e., the bending of light rays by gravitational fields), which, in the case of a black hole, exaggerates the size of the shadow. This can be seen by examining Fig.~\ref{fig:pathlengthening}: the black hole's gravitational field focuses the rays close to the unstable-photon region, so that two rays that were widely separated at the observer may both plunge. 

Another key consequence of the switch from a Newtonian to a relativistic picture is the introduction of the event horizon, the one-way causal membrane that separates the black hole's interior from its exterior, meaning that no information can be transmitted from the former to the latter. Although the horizon ($R=2GM/c^2$ in the Schwarzschild case) is a black sphere, it is actually the unstable-photon region (at $R=3GM/c^2$ in the Schwarzschild case), which perfectly captures all rays emanating from a distant observer that cross its boundary. It plays the role of the dark sphere we introduced in the Newtonian scenario. Crucially, that spheroidal region appears to us as sharp-edged; the reason for this is that whether or not a ray crosses this region is a yes/no question. Rays that intersect the unstable-photon region (and only those rays) are blocked from traveling further, thus reducing their path lengths, particularly through the emission region of a typical accretion disk (see the blue and red rays in Fig.~\ref{fig:pathlengthening}). Note that, as the black hole is engulfed by its emission region, the observed dip in intensity will never be complete, i.e., the shadow will never be perfectly dark. 

Because the rays are bent toward the singularity by the black hole's gravitational field, rays that `graze' the photon sphere begin to `loop around' it; in the extreme case of a photon that ends up exactly on the photon sphere, it will continue to do so indefinitely. We shall call this effect, which occurs for rays that pass near the boundary of the unstable-photon region, the `path lengthening effect'. For optically thin sources, the path length of a ray through the source is a proxy for the intensity picked up by that ray. Thus, the path-lengthening effect enhances the source's brightness around the edge of the shadow. We note that the path-lengthening effect occurs both inside of and outside of the shadow, i.e., for optically thin sources, the boundary of the black-hole shadow appears brighter to us than its center (note that this effect can also result from emissivity profiles that increase toward the center; see, e.g., Fig.~4 in \citet{narayan2019}). The blocking effect, however, breaks this apparent symmetry, and renders the interior of the shadow darker than its exterior.

Summing up, GR implies two significant changes, namely: i) The size of the shadow is increased due to gravitational lensing, and ii) The brightness of the cloud just outside the shadow (as well as just inside of it) is increased due to the path-lengthening effect.

\begin{figure}[htp]
\centering
\plotone{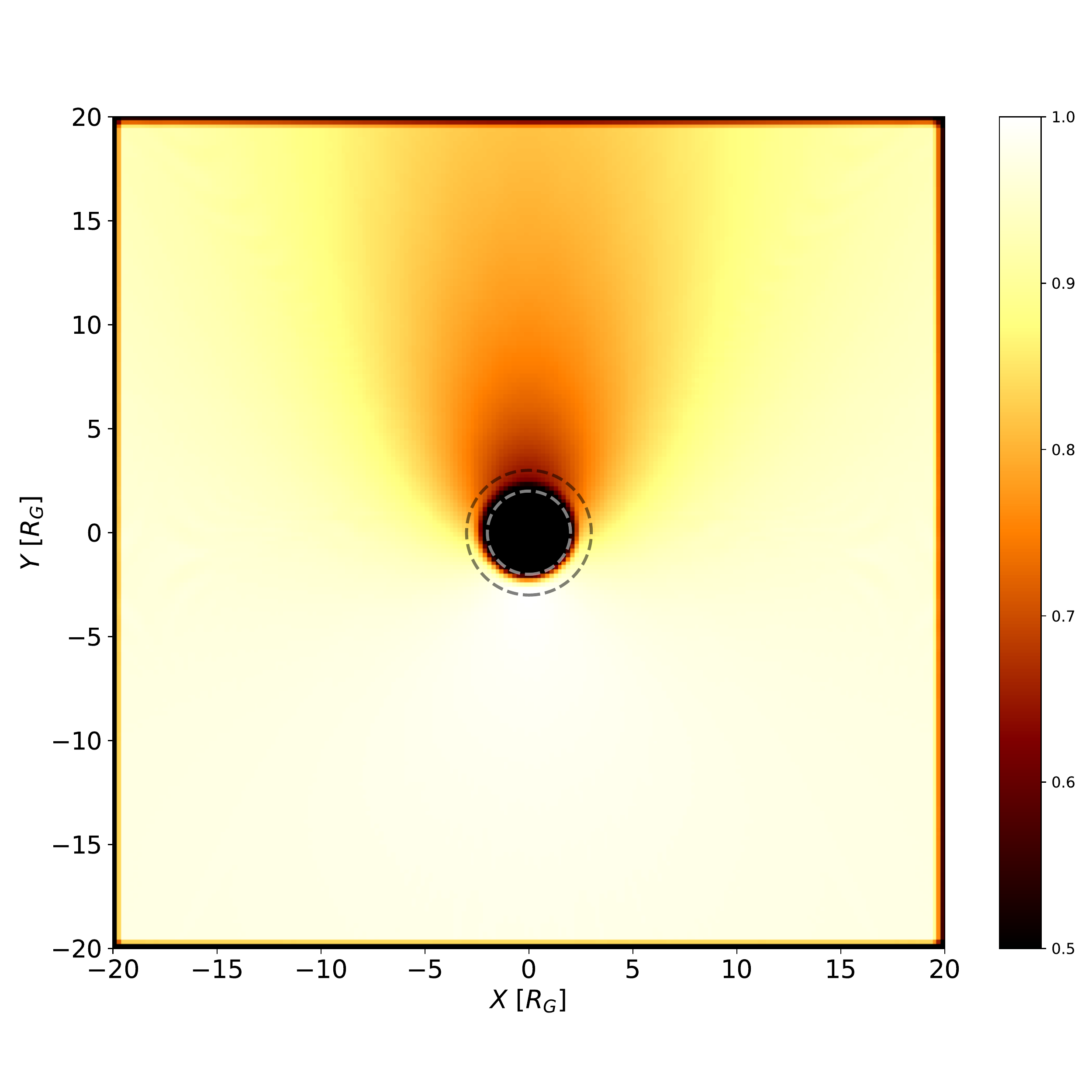}
\figcaption{Density map of null geodesics that reach a distant observer and arrive nearly parallel (below the bottom of the image), in the Schwarzschild spacetime. Unlike the Newtonian case shown in Fig.~\ref{fig:blockedregion}, flux may reach the observer from everywhere outside the event horizon (marked by the white dashed circle; the black dashed circle represents the unstable-photon region). However, a region is still observed from which lower-than-average flux reaches the observer. \label{fig:raydensityplot}}
\end{figure}

We emphasize that, while these two effects significantly alter the black-hole shadow - increasing the shadow's size and amplifying the source's brightness near its boundary -- the primary reason for why we see a shadow is still the blocking effect, i.e., the reduced intensity from rays that intersect the unstable-photon region. That remains true, even if Fig.~\ref{fig:raydensityplot} shows that in the case of GR, light can reach the observer from `behind' the black hole due to gravitational lensing. In the Newtonian case, the volume behind the absorbing object appears completely dark; in GR, it does not. However, the light from `behind' the black hole that reaches the observer, will appear to originate from outside the shadow to the observer, and only enhance the contrast of the shadow further. 

Finally, we note that in optically thick, geometrically thin emission regions, such as the Novikov-Thorne thin-disk model, neither the blocking effect nor the path-lengthening effect can play a role, since, for these models, a ray's path length inside the emitter is always zero.

\section{Radiative transfer: numerical calculations}
\label{sec:numerical}

Our next step is to produce images, using simple emission models, to investigate the effects described in the previous section in a realistic scenario. Since the appearance of an arbitrary emission region surrounding a black hole cannot generally be calculated in closed form, we turn to a numerical approximation. We employ the general-relativistic radiative transfer (GRRT) code {\tt RAPTOR} \citep{bronzwaer2018}. In this section, we will present image maps of several different source models, in order to investigate the behavior of the shadow.

\subsection{Spherical emitters}

\begin{figure*}[htp]
\centering
\gridline{\fig{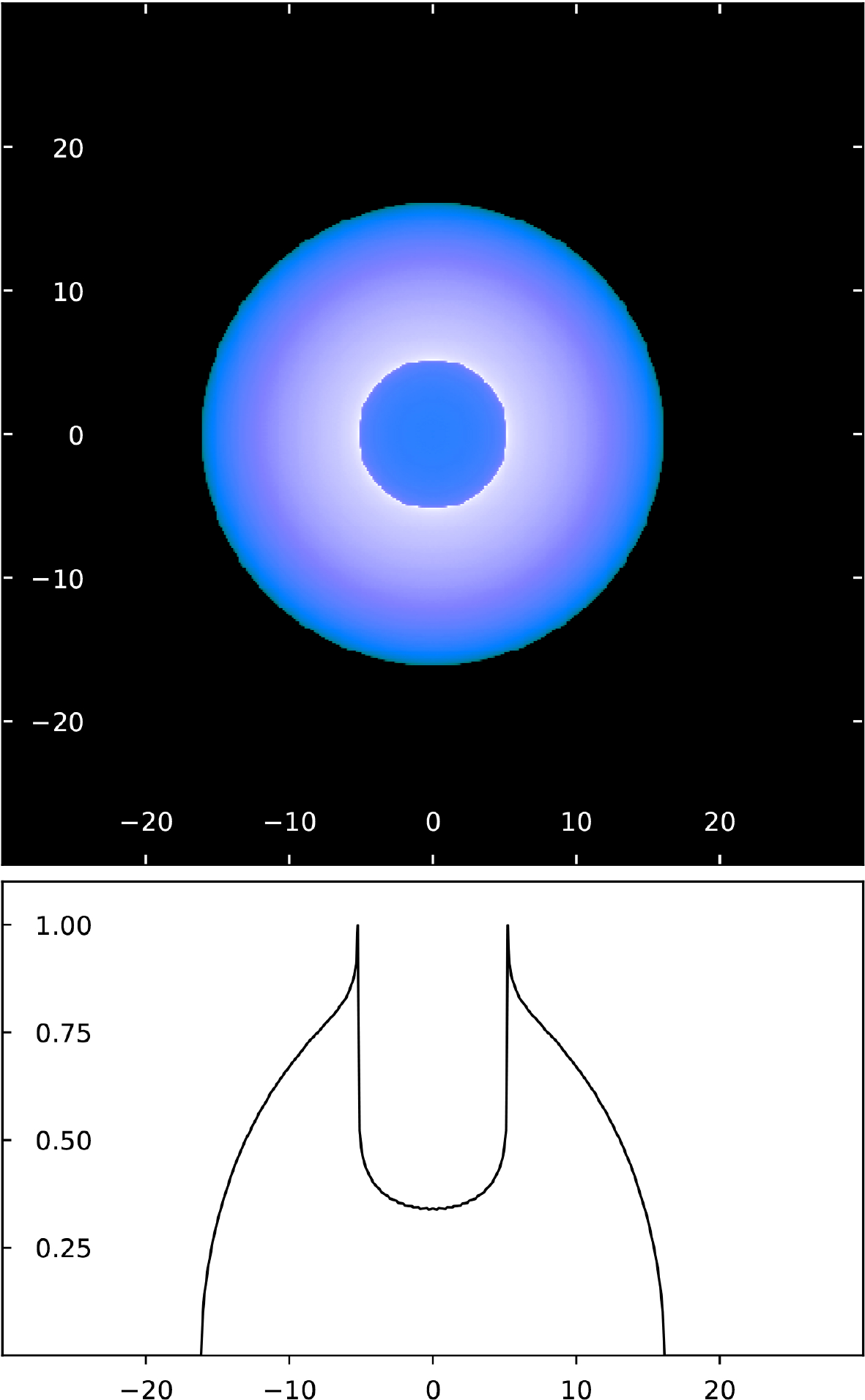}{0.42\textwidth}{}
          \fig{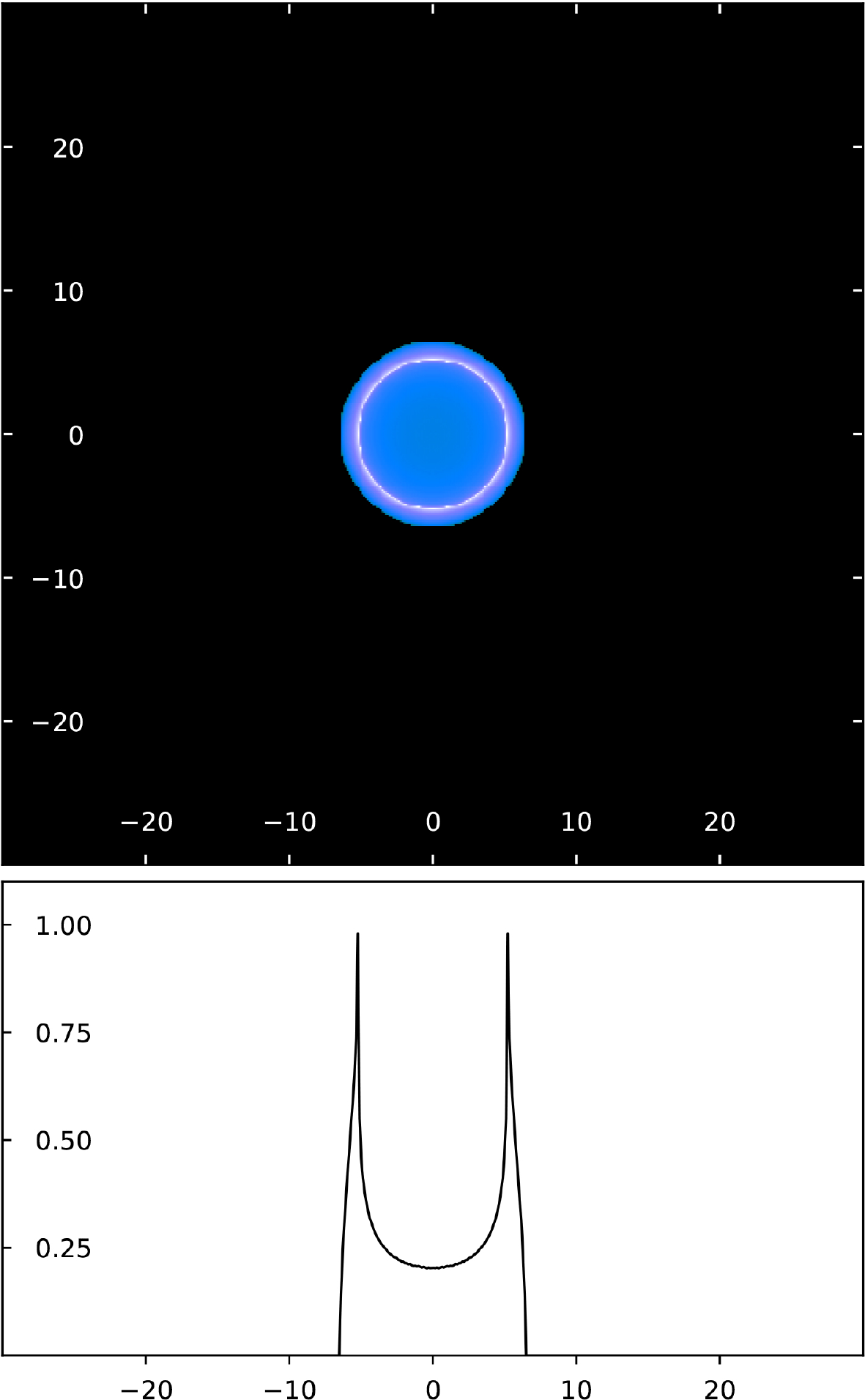}{0.42\textwidth}{}}
\figcaption{Upper panels: maps of null-geodesic path length inside a spherical volume of radius 15 $R_{\rm G}$ (left) and 5 $R_{\rm G}$ (right). The axes of the upper panels are measured in $R_{\rm G}$. Lower panels: central, horizontal cross-sections of the upper panels; the horizontal axis represents the null geodesics' impact parameter, while the vertical axis represents the normalized path length. Note the step-like dip in intensity that occurs at the boundary of the shadow. \label{fig:Rout20}}
\end{figure*}

Figure \ref{fig:pathlengthening} shows the geometry of null geodesics for a distant observer looking at a Schwarzschild black hole. In order for the observer to see anything, an emission region must be present that intersects at least some of the viewing rays. The simplest possible case is that of a spherical emitter that is completely optically thin and homogeneous. In this case, the path length of the section of the ray that is inside the source is a proxy for the brightness received by the corresponding camera pixel. Figure \ref{fig:Rout20} shows normalized image maps of the path length taken through the source, for two spherical volumes (emitter proxies) of radii 15 and 5 $R_{\rm G}$. Recalling our discussion in Section \ref{sec:theory}, we see that, for the larger emitter, the darkening factor is roughly $1/2$, as in the Newtonian case; for the smaller emitter, the path-lengthening effect dominates, instead.

\begin{figure}[htp]
\centering
\plotone{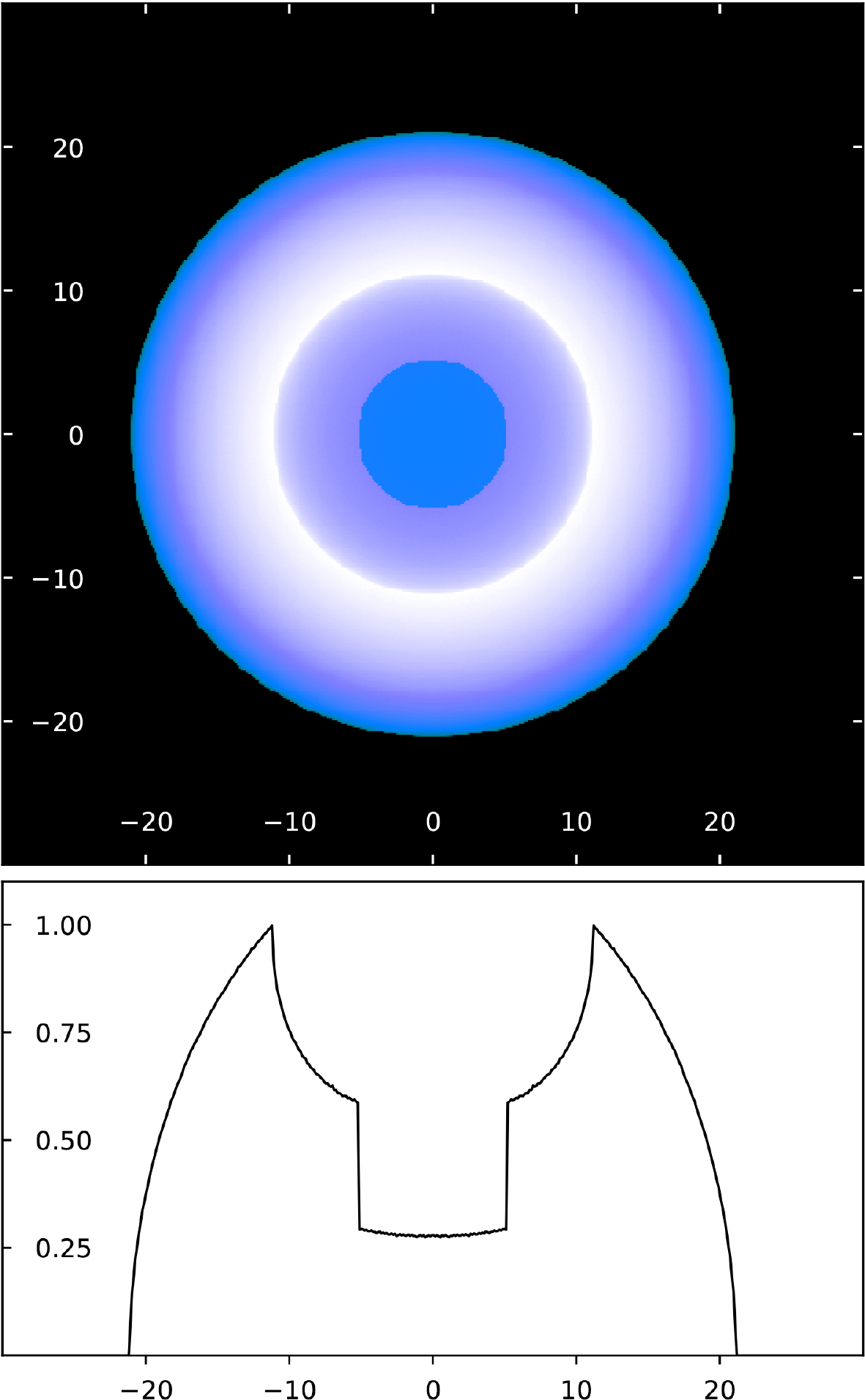}
\figcaption{As Fig.~\ref{fig:Rout20}, but with an outer volume radius of 20 $R_{\rm G}$ and an inner volume radius of 10 $R_{\rm G}$ (i.e., no emission originates from within the inner 10 $R_{\rm G}$). \label{fig:pathlength_innercutout}}
\end{figure}

Figure \ref{fig:pathlength_innercutout} shows a similar path-length map, for a volume with outer radius 20 $R_{\rm G}$ and inner radius 10 $R_{\rm G}$ (so that the inner 10 $R_{\rm G}$ are evacuated). In this case, no path-lengthening effect is observed, because the region in which that effect occurs (near the boundary of the unstable-photon region) is evacuated. Thus, we see that in cases like Fig.~\ref{fig:pathlength_innercutout}, the blocking effect dominates while the path-lengthening effect (and, correspondingly, the `photon ring') vanish; in other cases, such as the right panel of Fig.~\ref{fig:Rout20}, it is the path-lengthening effect that dominates. We note that this is completely analogous to the models depicted in Fig.~4 of \citet{narayan2019}, which consists of a spherical accretion flow with a spherical, inner `cutout' of varying radii; in this model, too, the shadow remains constant, even when the cutout radius is varied, and the photon ring becomes invisible. This provides a counterexample to the claim by \citet{gralla2020} that the photon ring is generic (and that the shadow is not).

\begin{figure}[htp]
\centering
\plotone{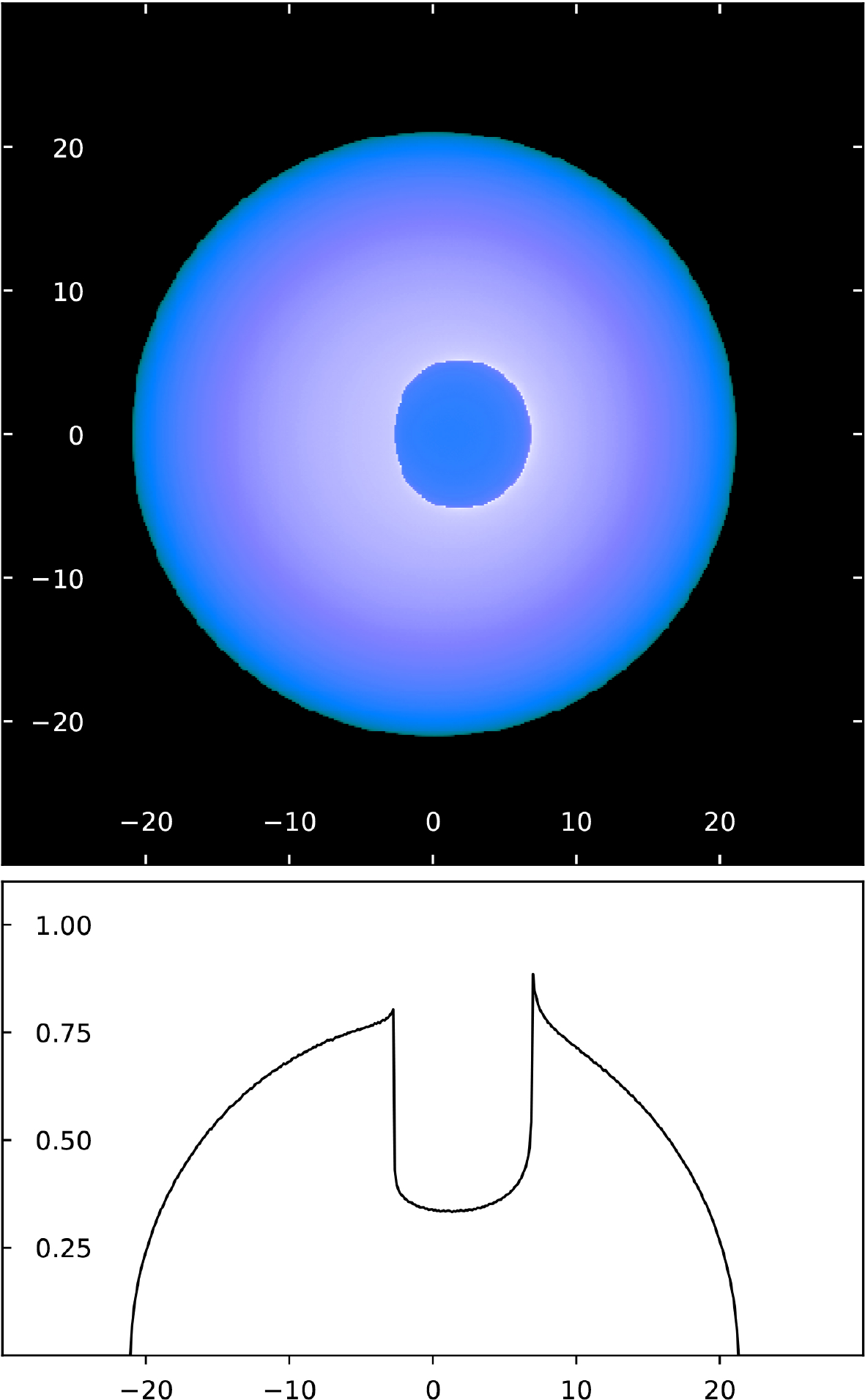}
\figcaption{As Fig.~\ref{fig:Rout20}, but for a Kerr spacetime with $a=0.9375$ and an outer volume radius of 20 $R_{\rm G}$. \label{fig:pathlength_highspin}}
\end{figure}

Figure \ref{fig:pathlength_highspin} shows a path-length map for a volume with external radius 20 $R_{\rm G}$, but for a Kerr spacetime with $a=0.9375$. In this case, the non-zero spin gives rise to frame-dragging, and the characteristic asymmetric appearance of the black-hole shadow. Note, in particular, how the path-lengthening effect is suppressed on the image's left-hand side. This is because rays that are coming toward the observer on this side of the black hole receive a `boost' due to frame dragging - these rays do not plunge, but are quickly `swept clear' of the unstable-photon region due to the rotation of spacetime. 

\subsection{Toroidal and cylindrical emitters}
\label{sec:torus}

In order to understand the effects on the shadow of breaking spherical symmetry, as well as the influence of the emission-region geometry (i.e., the astrophysical circumstances), we now turn to toroidal and cylindrical emitters. These models break spherical symmetry because they carry a non-zero angular momentum, making them more realistic representations of astrophysical emission regions. Such emitters are geometrically thick and optically thin, corresponding to radiatively inefficient, low-density, high temperature emission regions, which occur around low-luminosity AGN in accretion flows and jets \citep{falcke2001}. Because the accretion flow carries angular momentum, the effect of relativistic boosting becomes apparent for these models, increasing the intensity of the side of the torus (or cylinder) on which the flow moves toward the observer and diminishing the intensity of the receding side (the effect vanishes when the emitter is seen face-on, and is strongest when the observer is in the black hole's equatorial plane). It is interesting to note that, despite the absence of the path-lengthening effect on that side of the black hole (see Fig.~\ref{fig:pathlength_highspin}), this region is still the brightest region in an image of a rotating black hole, due to the relativistic beaming of the accretion flow.

In our analytical model, the accretion flow's local velocity vector is constrained to be tangential to a circular path. The velocity is then scaled as follows:
\begin{equation}
\Omega = \frac{d\phi}{dt} = \frac{1}{\sqrt{r^{3/2} + a}},
\label{eq:velvec}
\end{equation}
which, in combination with the demand that the vector's norm equals -1 and a choice of rotation direction (prograde with respect to the black hole's sense of rotation in our case), determines the vector elements. Only in a section of the equatorial plane do the paths implied by these tangent vectors correspond to timelike geodesics (the stable, circular orbits just outside of the black hole's innermost stable circular orbit (ISCO)). Thus, other regions of the flow must be affected by pressure gradients or magnetic fields.

The emission coefficient of our toroidal model is proportional to a Gaussian profile defined as follows:
\begin{equation}
    j_{\nu}\left( {\bf r} \right)= \exp{\left[-\frac{\left( {\bf r} - {\bf r_s} \right)^2}{\sigma^2}\right]},
\end{equation}
where ${\bf r_s}$ is the location of the torus center in the $r, \theta$ plane (since it is in the equatorial plane, $\theta_s=\pi/2$), and $\sigma$ is the parameter that sets the torus' minor radius.

Using this model, we now plot several intensity maps for qualitatively different scenarios, to illustrate how the different physical processes behind the black-hole shadow contribute.

\begin{figure}[htp]
\centering
\plotone{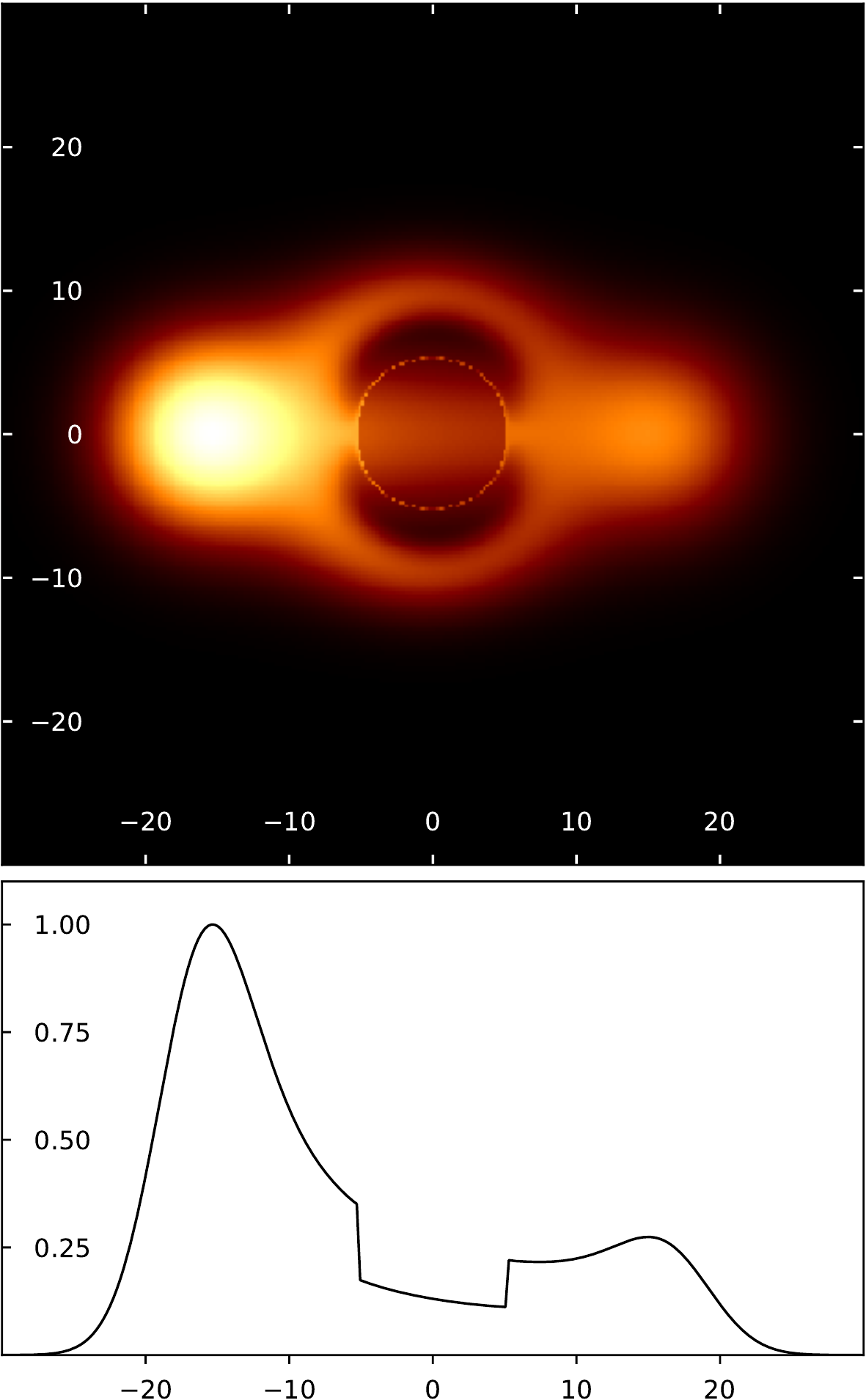}
\figcaption{Intensity map of our toroidal emission model with $R_{\rm maj} = 16$ $R_{\rm G}$ and $R_{\rm min}=3$ $R_{\rm G}$, in a Schwarzschild spacetime, for a distant observer situated in the black hole's equatorial plane.  \label{fig:torus_evacuated}}
\end{figure}

\begin{figure}[htp]
\centering
\plotone{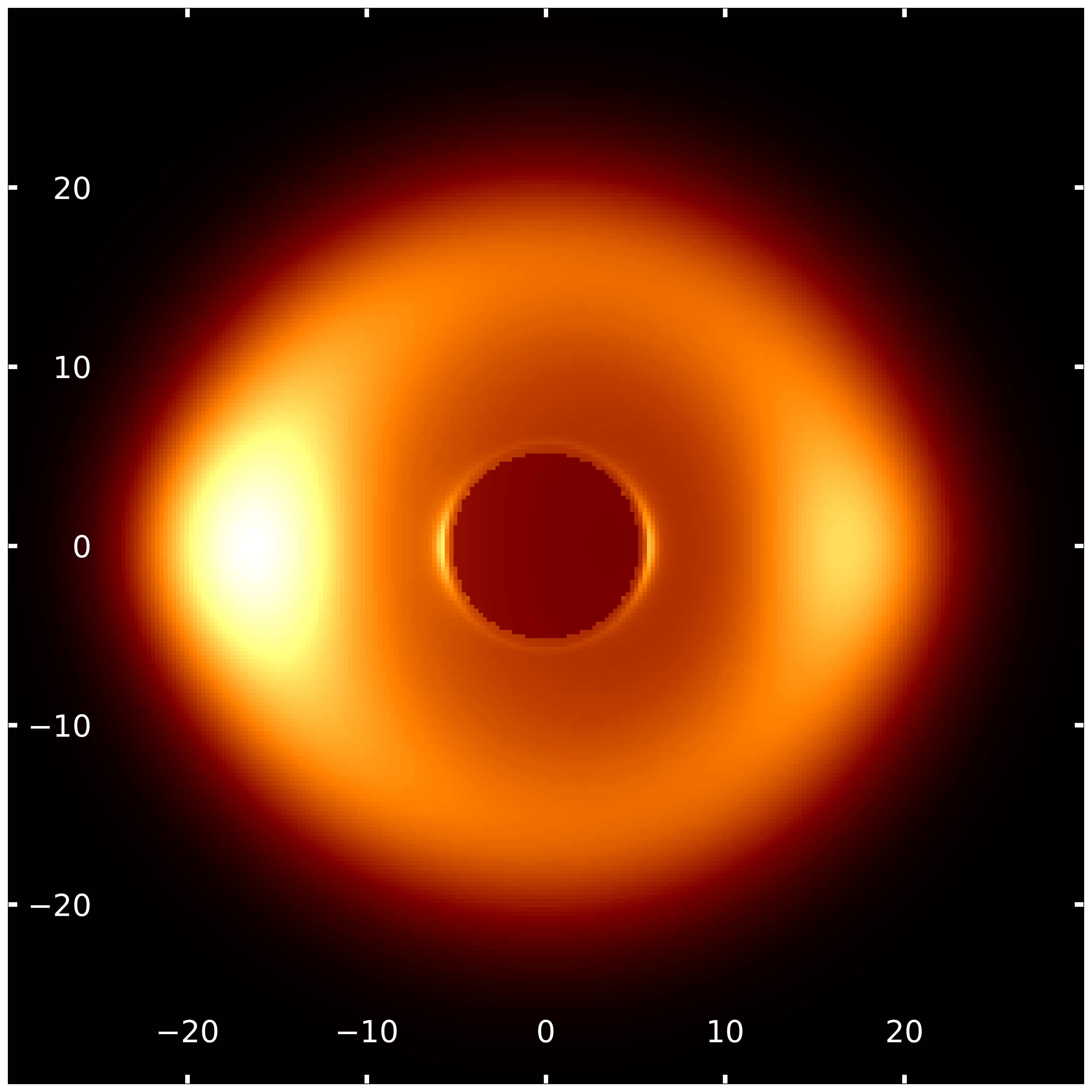}
\figcaption{The model shown in Fig.~\ref{fig:torus_evacuated}, revolved around the image X-axis (a summation of intensity maps for all inclination angles). \label{fig:summing}}
\end{figure}

Figure \ref{fig:torus_evacuated} shows a torus with major radius 16 $R_{\rm G}$ and minor radius 3 $R_{\rm G}$, viewed from the equatorial plane. In this model, the emission region lies well outside the black hole's ISCO, and demonstrates the `evacuation' effect discussed in \citet{bronzwaer2020}, causing an enlarged CBD around the black hole's true shadow, although the sharp-edged optical signature at the shadow remains. Figure \ref{fig:summing} shows the effect of revolving this emission region around the image X-axis, i.e., summing the images from all inclination angles; this effectively `fills in' the evacuated regions, but the step-like dip in brightness at the shadow remains constant, thus producing a clearly observable shadow. 

\begin{figure}[htp]
\centering
\plotone{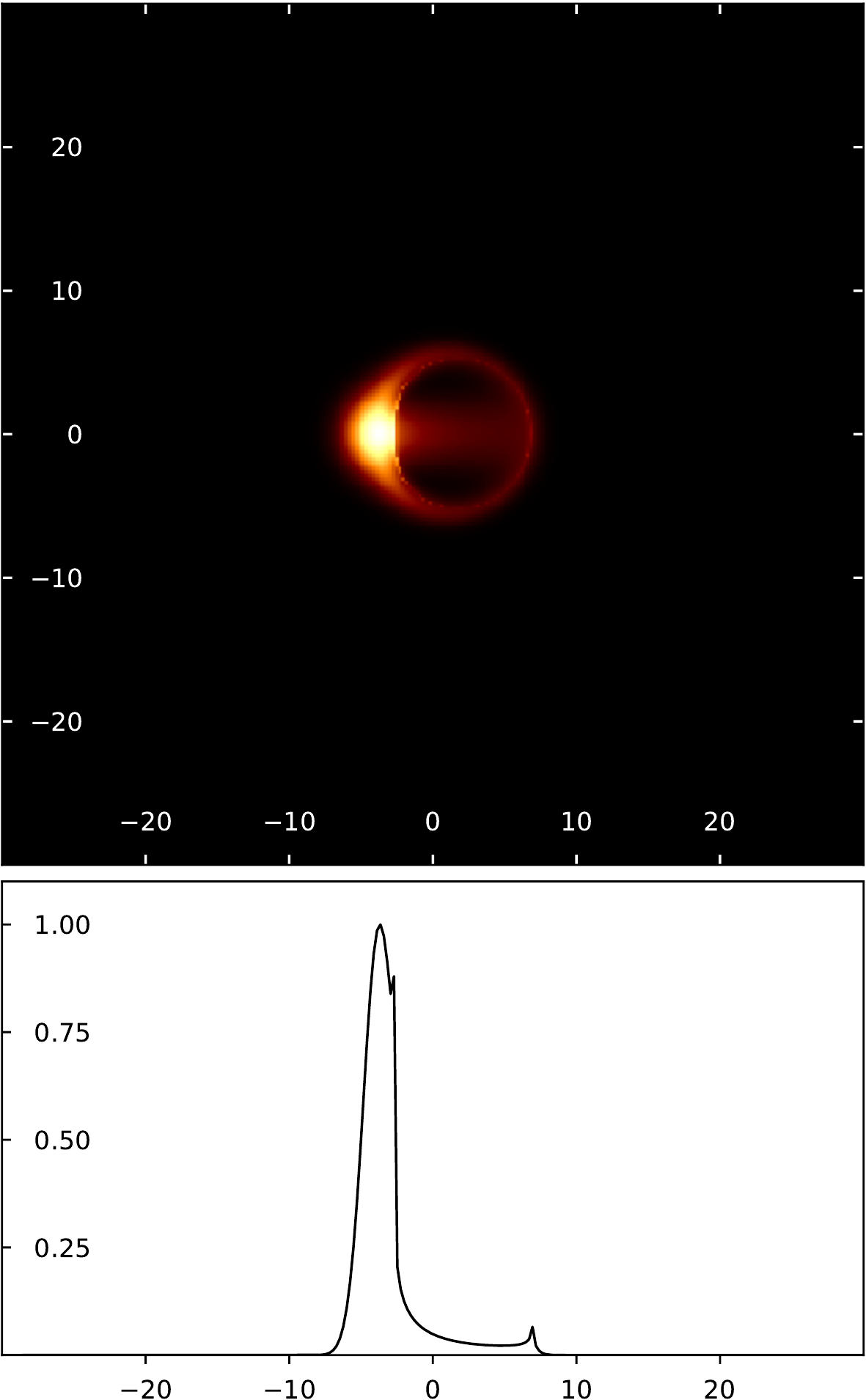}
\figcaption{Intensity map of our toroidal emission model with $R_{\rm maj} = R_{\rm ISCO} +1$ $R_{\rm G}$ and $R_{\rm min} = 1$ $R_{\rm G}$, in a Kerr spacetime with $a=0.9375$, for a distant observer situated in the black hole's equatorial plane.  \label{fig:torus_highspin}}
\end{figure}

Figure \ref{fig:torus_highspin} shows a Kerr black hole ($a=0.9375$) surrounded by a toroidal flow with $R_{\rm min} = 1$ $R_{\rm G}$ on an orbit that lies 1 $R_{\rm G}$ outside of the ISCO. Note that the high spin causes a strong relativistic boosting effect and thus a small emitter size, as we see only the boosted region. Compared to the Schwarzschild case, the ISCO lies much closer to the unstable-photon region for a high-spin black hole, and thus, the evacuation effect is reduced, and a clear black-hole shadow is observed for co-rotating disks around high-spin black holes. Interestingly, Figs.~\ref{fig:torus_evacuated} and \ref{fig:torus_highspin} are rather similar in appearance to the equivalent SANE disk models discussed in \citet{bronzwaer2020}.

\begin{figure*}[htp]
\centering
\gridline{\fig{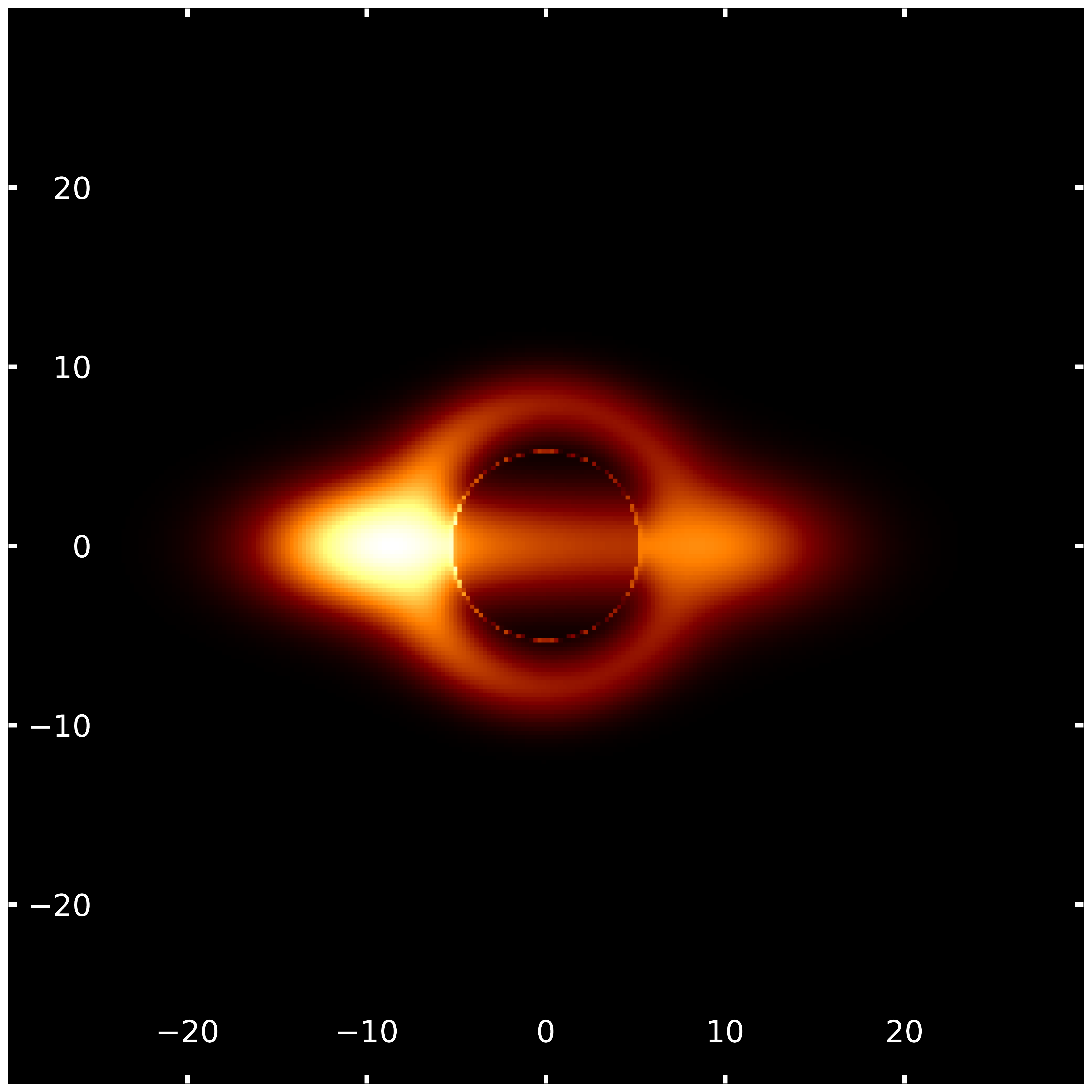}{0.32\textwidth}{}
          \fig{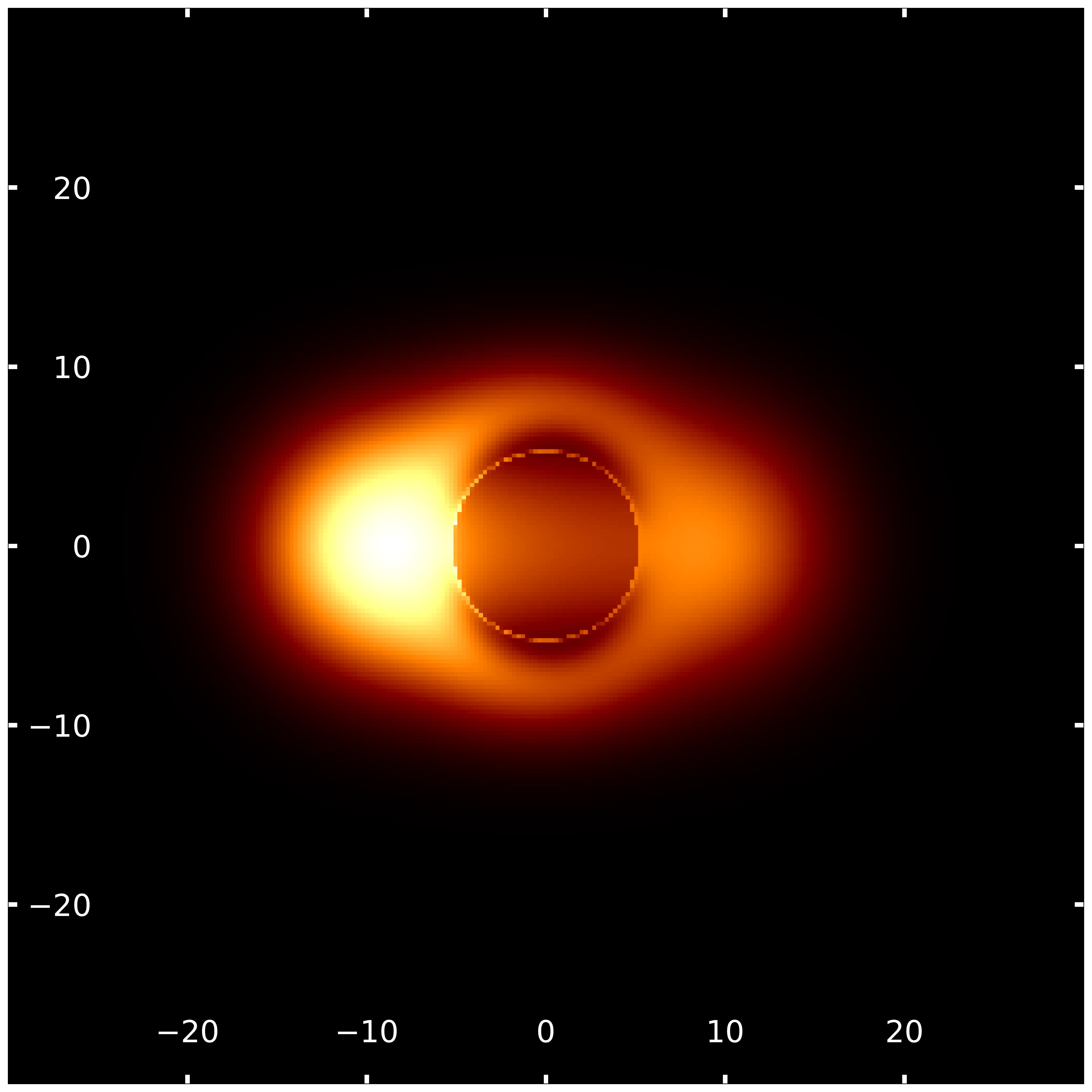}{0.32\textwidth}{}
          \fig{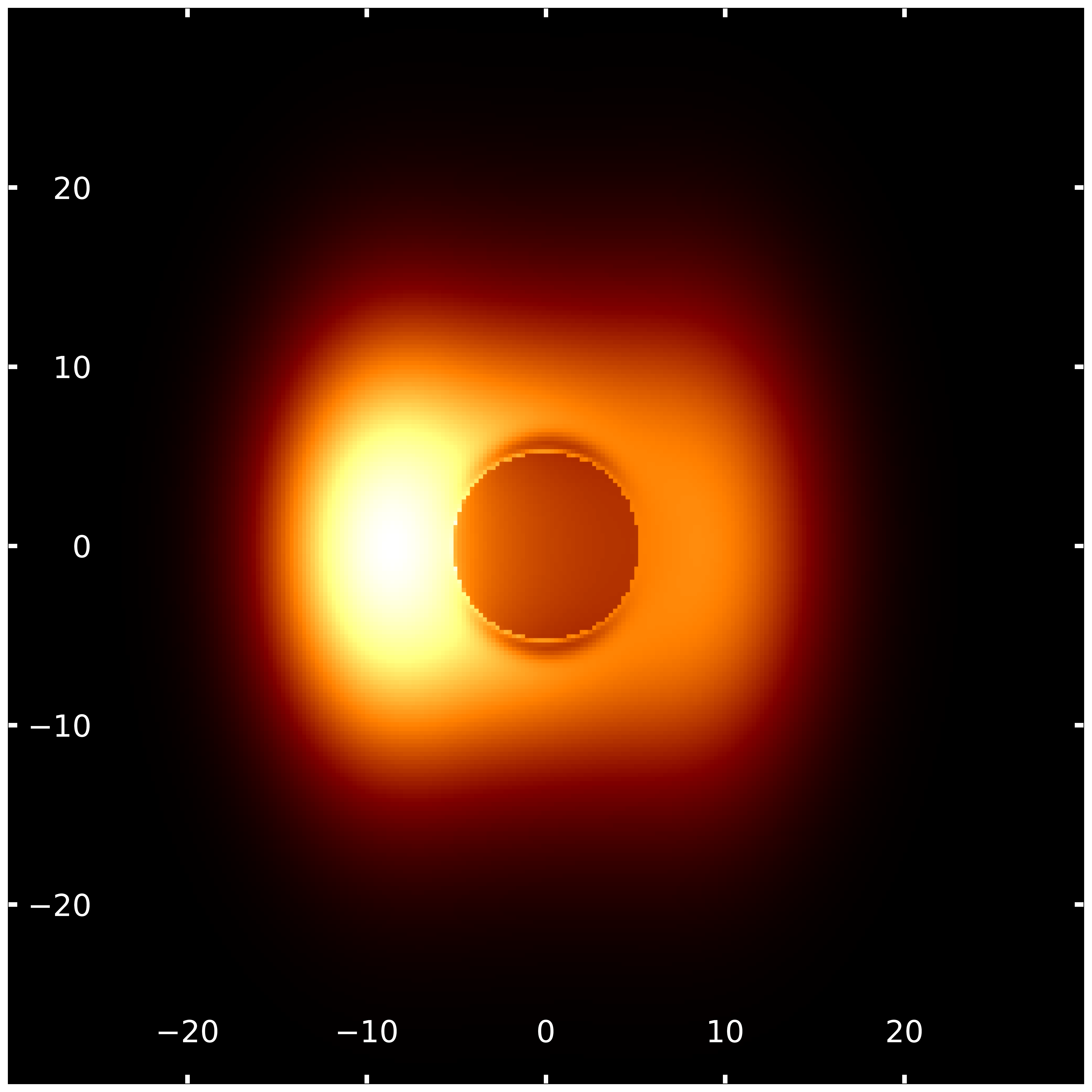}{0.32\textwidth}{}}
\figcaption{Intensity maps for a toroidal emitter with $R_{\rm maj} = 9$ and $R_{\rm min} = 3$, with a $z$-scaling factor of 0.5 (left panel), 1 (center panel), and 2 (right panel). \label{fig:fattening}}
\end{figure*}

In order to explore the range of possible models between the geometrically thin and geometrically thick regimes, Fig.~\ref{fig:fattening} shows the transition from a thin to a thick torus, illustrating the decreasing severity of the evacuation effect for thinner torii as the disk gets thicker. For very thick disks, a clear shadow can be seen, reminiscent of the spherical case, as the blocking effect is on full display. Thinner emission regions, on the other hand, contain relatively more evacuated regions, interfering with the blocking effect and rendering the shadow less easily observable; a step-like dip in brightness is still observed, but the CBD has a more complicated shape.

From the preceding examples, we see that in the case of toroidal emitters, the visual signature (step-like dip in brightness) that is coincident with the BHS arises due to a complex interplay of gravitational lensing (causing its exaggerated size), the path-lengthening effect (increase of the brightness along its edge), and the blocking effect (relative darkening of rays that intersect the unstable-photon region). It is remarkable that these effects all contribute to a visible signature that lies on the edge of the shadow (as seen by a distant observer). For a more detailed look at GRMHD-based toroidal emission models, see \citep{bronzwaer2020}.

\begin{figure}[htp]
\centering
\plotone{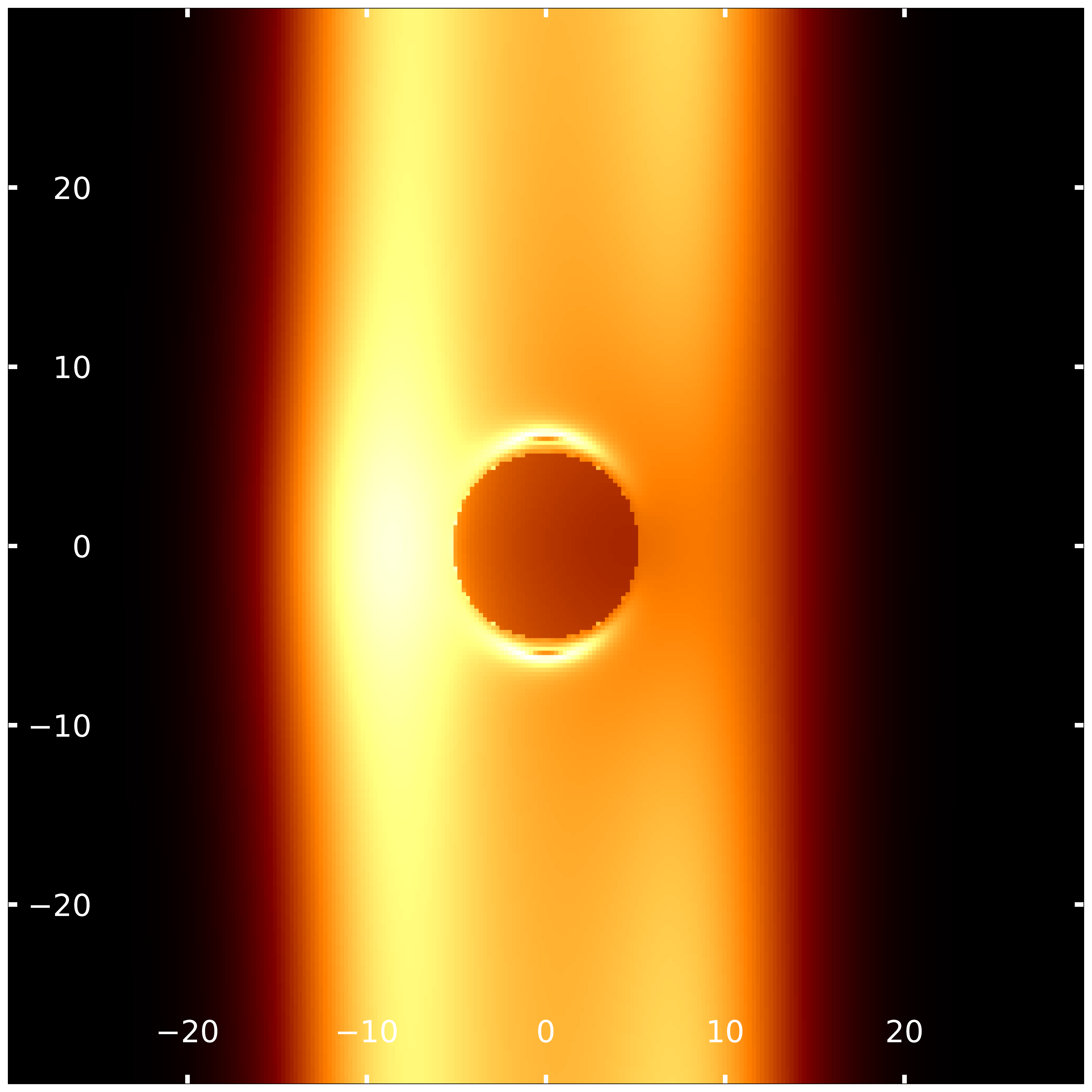}
\figcaption{Intensity map of our cylindrical emission model with a radius of 9 $R_{\rm G}$ and a wall thickness of 6 $R_{\rm G}$ (so that the inner core is evacuated). \label{fig:cylinder}}
\end{figure}

Figure \ref{fig:cylinder} shows our cylindrical emission model, which is a simple model for a disk-jet accretion flow. The cylinder has a hollow core; its radius is 9 $R_{\rm G}$ and its wall thickness is 6 $R_{\rm G}$. As in the toroidal case, the cylinder is given a purely azimuthal velocity vector which is scaled by Eq.~\ref{eq:velvec}. Lensed images of the emission region's hollow inner core can be observed just above and below the shadow. Note, however, that the step-like dip in intensity at the BHS is visible in each case; this model contains only tiny and well separated evacuated regions, compared to the toroidal models, so that essentially a complete shadow is observed. 

\subsection{Thin-disk emitters}

Emission regions with a higher density and radiative efficiency and a lower temperature than the RIAF flows discussed in the preceding section may form in, e.g., black hole binaries and QSO's. A model was developed to describe such flows (\citealt{shakura1973}; \citealt{novikovthorne1973}); it consists of a geometrically thin, optically thick disk confined to the equatorial plane and terminating abruptly at the ISCO. The velocity vector is identical to the previous section, so that disk elements move along circular, timelike geodesics. As the disk is geometrically thin and optically thick, its associated emission coefficient is a delta function along a ray, and the ray terminates upon encountering the disk. 

\begin{figure*}[htp]
\centering
\gridline{\fig{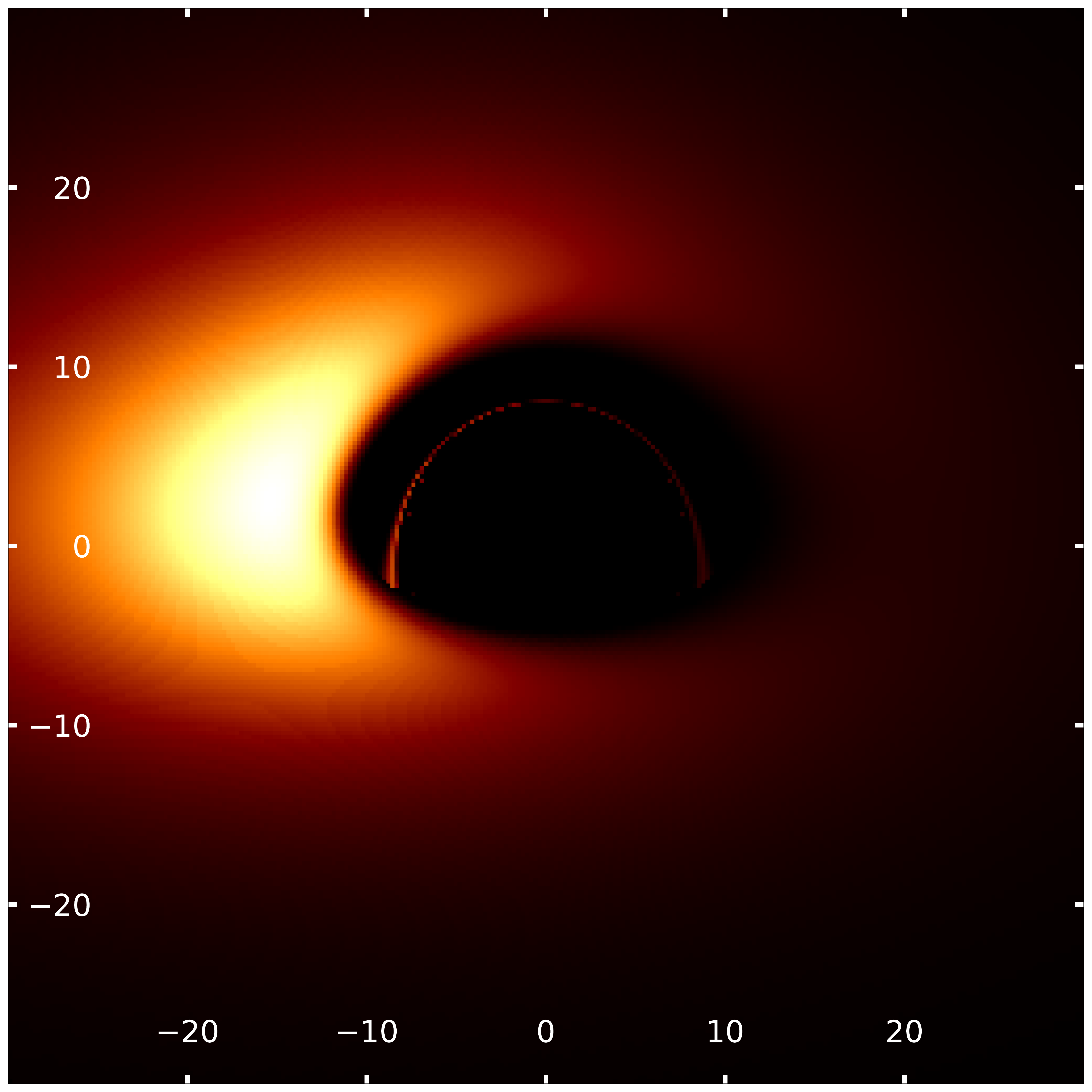}{0.32\textwidth}{}
          \fig{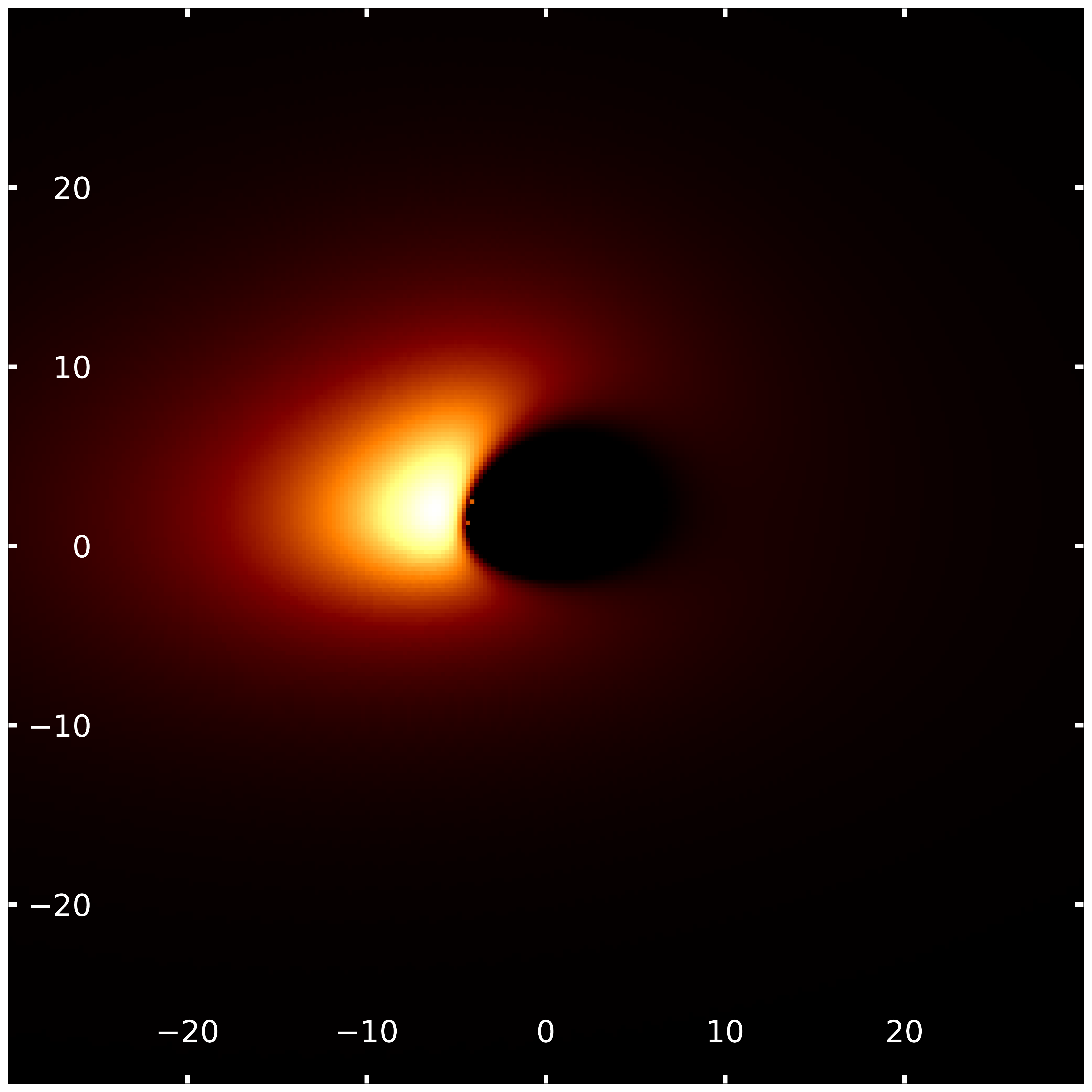}{0.32\textwidth}{}
          \fig{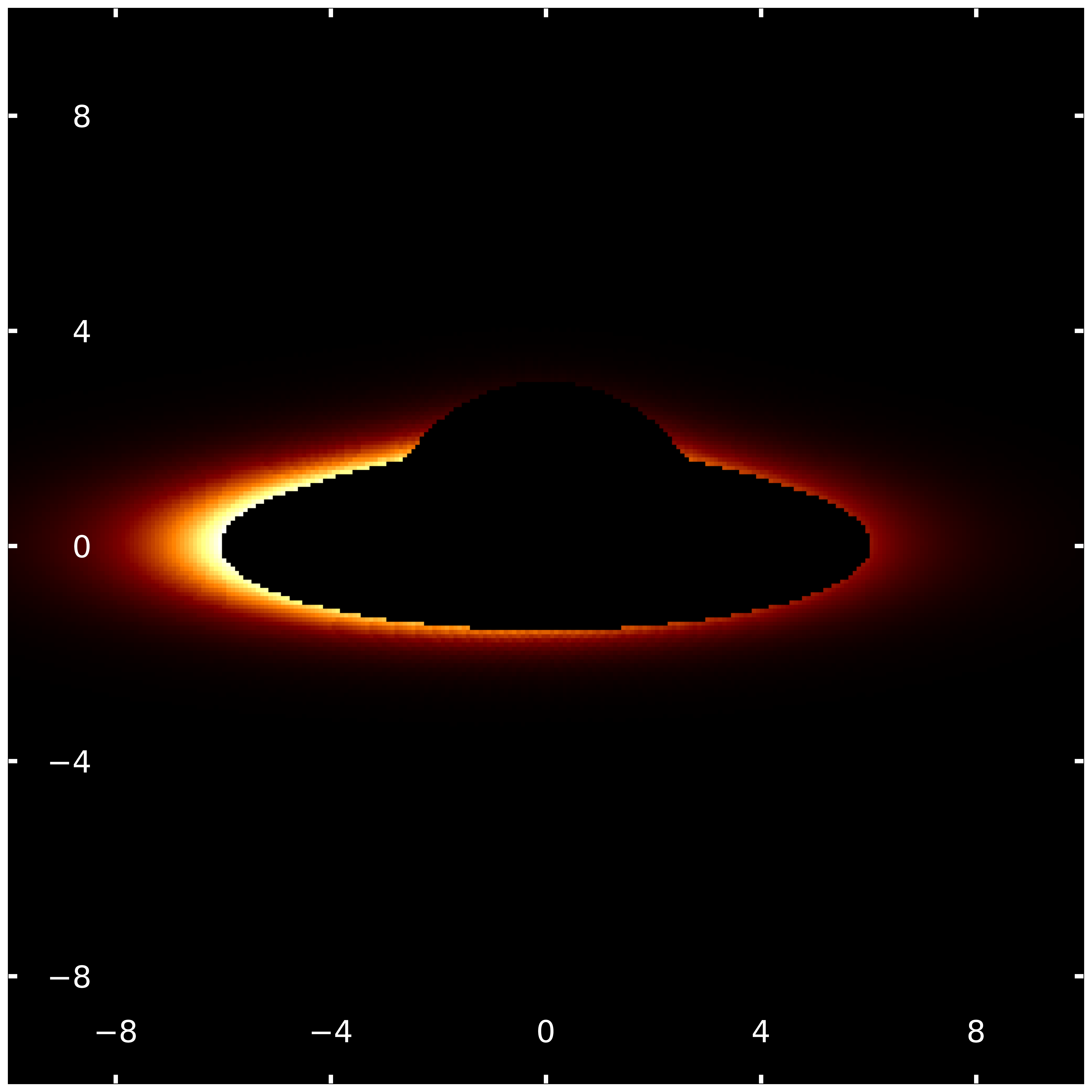}{0.32\textwidth}{}}
\figcaption{Intensity maps of the thin-disk model by \citet{novikovthorne1973}, for an observer-inclination angle of $75 \deg$, and a black-hole spin of $a = 0$ (left panel), and $a = 0.9375$ (center panel). Right panel: Newtonian analog of thin-disk model, also viewed at $i=75 \deg$.  \label{fig:thindisk}}
\end{figure*}

Figure \ref{fig:thindisk} shows intensity maps of this model for an observer-inclination angle of $75 \deg$ and two different black-hole spins, as well as a Newtonian analog of the model, which is included for comparison; note the different length scale of the Newtonian image, which indicates the difference in apparent size of the source due to the absence of gravitational lensing. Regarding the BHS for these models, as in the case of toroidal emitters, the evacuation and obscuration effects are apparent (the evacuation effect is rather prominent for this model, as we exclude any emission from within the ISCO, a crude assumption which makes it a sort of worst-case scenario). 

In this model, the blocking effect and the path-lengthening effect are entirely absent, as there is no geometrically thick, optically thin flow through which to measure path length in the first place. Thus, no step-like drop in brightness occurs at the shadow, and the appearance of the CBD is much more model-dependent than in the case of an optically thin, geometrically thick emission region. However, the gravitational lens effect still displaces the emission coming from behind the unstable-photon region (with respect to the observer) so that it appears to come from outside of the unstable-photon region. Thus, as the disk's inner radius approaches the unstable-photon region, a CBD may be observed that is nearly coincident with the BHS. This occurs at high spins, for which the ISCO and the unstable-photon region both shrink down, approaching the size of the event horizon for $a \rightarrow 1$ \citep{bardeen1972}. At low spins, or particularly retrograde spins, the ISCO is pushed further out, and the evacuation effect described in \citet{bronzwaer2020} exaggerates the size of the observed CBD. The obscuration effect described in that same paper also occurs in the thin-disk model, for most inclinations. However, the evacuated region differs from the true shadow's interior, because the celestial sphere is visible in the former, and, depending on the observing frequencies, astrophysical objects (e.g., distant AGN) will be visible there, thus rendering the two regions distinguishable, at least in theory. Of course, in practice, the brightness of the lensed objects that appear in the evacuated region is much lower than that of the accretion disk, and it may also be difficult to disentangle this emission from any foreground emission. 


\section{Discussion and Conclusions}
\label{sec:conclusions}

A black hole's shadow is the projection of a its unstable-photon region on an observer's sky. If the black hole is surrounded by an optically thin, geometrically thick emission region, a step-like dip in brightness will be observed that is coincident with the shadow. This dip may or may not be coincident with the overall CBD that is observed (which may itself have a complicated structure, see, e.g., \citet{chael2021}). The purpose of this work was to get a more intuitive understanding of the shadow as an astrophysical observable. 


In the case of spherical emission regions, the primary mechanism behind the black hole shadow is the blocking effect; the shortening of path lengths for null geodesics that intersect the photon sphere and terminate at the event horizon, versus null geodesics that do not intersect the photon sphere.
For these spherical models, the shadow may be readily identified by an edge-detection scheme \citep{PsaltisOzelChan2015a} or some form of crescent shape fitting \citep{KamruddinDexter2013a,EHT2019VI}.

Besides the blocking effect, the path-lengthening effect occurs, whose observational signature is a relative enhancement of brightness at the edge of the black-hole shadow. We note that this effect occurs both inside and outside of the unstable-photon region (and thus the shadow); the cusp-like increase in brightness caused by the path-lengthening effect peaks on the shadow's boundary, and decays in both the outward and inward directions. However, when the observable effects of the blocking effect and the path-lengthening effects are combined, the part of the cusp that is interior to the black-hole shadow appears reduced in brightness (see the profile of the left panel of Fig.~\ref{fig:Rout20}), so that the rays that travel exterior to the unstable-photon region dominate the image, and the shadow's interior is overall diminished in brightness.
These strongly bent light rays, which travel near the edge of the unstable-photon region (but which are outside of it), produce a complex system of lensed images of the black hole's environment. This is often called the photon or light ring, and it has been invoked for tests of gravity \citep{luminet1979,FalckeMeliaAgol2000b,JohannsenPsaltis2010a,JohnsonLupsascaStrominger2020a,GrallaLupsascaMarrone2020r}. The photon ring contains multiple projections of the surrounding emission region, and approaches the shadow asymptotically the more often a light ray goes around the black hole (thus approaching the peak of the cusp); \citet{MoscibrodzkaFalckeShiokawa2016a} show that, in the case of M87*, the lensed, first-order image of the counter-jet can actually dominate the image of the source (see also Appendix A of \citet{davelaar2019}). We have shown that a step-like dip in brightness at the shadow, and even a CBD that is coincident with the shadow, can arise independently from the path-lengthening effect (and thus the photon ring). This will happen, for example, if a volume outside the black hole's ISCO is evacuated, as in Fig. 4 of \citet{narayan2019} and Fig.~\ref{fig:pathlength_innercutout} of this work. In certain models, the CBD can be `multi-tiered'; a step-like dip may occur at the shadow, along with an `inner shadow', which is a smaller, smooth-edged dip that occurs at the projection on the observer's sky of the intersection of the black hole's event horizon with its equatorial plane \citep{chael2021}.

Neither the blocking effect nor the path-lengthening effect is relevant with regard to the appearance of geometrically thin emission disks.  The dark region (CBD) inside of a thin disk can be larger or smaller than the BHS, as one observes a combination of the shadow and the evacuated region in the emission disk. This ambiguity disappears when the emission regions transits from a planar to a spherical structure (See Fig.~\ref{fig:fattening}). 

We note that the appearance of a sharp-edged CBD that is coincident with the projection of a dark object (such as a black hole's unstable-photon region) is itself not unique to GR. A perfectly light-absorbing object in flat (Newtonian) spacetime and surrounded by an optically thin emission region would produce a similar CBD. The everyday example we gave is that of a kettle in a campfire, or of a wick in a candle. However, the shadow in GR is exaggerated in apparent size by a factor of about 2.5 due to gravitational lensing. 
Moreover, the image can exhibit a bright photon ring due to path lengthening in the emission region, if the emission region (nearly) intersects the photon sphere. Both effects are due to light bending. The darkening itself, however, implies either the presence of an event horizon - a perfectly absorbing surface with zero reflectivity at all wavelengths - or of another type of surface which must nevertheless lie close to the event horizon, and thus is hidden from outside observers due to the strong gravitational redshift. This is in contrast to the everyday Newtonian example of the kettle in a campfire, where the initially dark kettle would heat up and eventually be seen in emission. The absence of such thermal emission has been used to infer the presence of a horizon in stellar-mass black holes \citep{NarayanGarciaMcClintock1997} and in M87* \citep{EHT2019VI}.

In observations, the darkness of the observed CBD is limited by the dynamic range and resolution of the observing instrument, as well as the accuracy with which foreground emission can be subtracted. Therefore, the presence or absence of a black hole's event horizon can only be verified to within certain limits. However, since the emission region is close to the black hole, any light-emitting matter moves at a significant fraction of the speed of light, whether falling into the event horizon, escaping in a plasma jet or wind, or rotating on bound
orbits. The dynamical time scale for these processes range from milliseconds for stellar-mass black holes to a couple of days for a supermassive black hole like M87*. Hence, integrated and averaged over human time scales, any irregularities in the foreground emission should be smoothed out.

In case the black hole is not completely surrounded by an emission region, an evacuated region \citep{bronzwaer2020} may be observed around the shadow. However, unlike the shadow itself, the darkness in these patches is of a fundamentally different nature. It will not be completely dark and smooth, but will contain a static lensed image of the universe. The same applies for the criticism of \citet{bardeen1974} by \citet{gralla2019}, who states that a black hole in front of a uniformly radiating surface, e.g. a stellar disk, appears larger than the projection on the observer's sky of the unstable-photon region. However, in this case, too, the `dark' region surrounding the true shadow is of a fundamentally different nature than the shadow itself, as it, too, is filled with emission from lensed images of distant sources in the entire universe. Unlike any rapidly moving foreground emission, this background image will remain static, although it will also be extremely faint.

Some compact objects without horizons (but with highly redshifted surfaces deep within their photon spheres) can also show shadows. However, such objects may be distinguishable from black holes through either re-radiation \citep{BroderickNarayan2006a} or reflection of emission. For example, in the absence of a photon-capture cross section, the central dark region in the boson star case \citep{VincentMelianiGrandclement2016a} is simply a lensed image of the star's central low-density region \citep{olivares2020}. Light could pass through the boson star itself. In rotating boson stars, plunging geodesics can become chaotic, and are reflected outwards towards infinity \citep{cunha2016, olivares2020}. One would expect to see a highly distorted, lensed image of the universe in lieu of a perfectly dark shadow, for such objects. Naked singularities could, in principle, also show a shadow \citep{ShaikhKocherlakotaNarayan2019a} as they are matched to a Schwarzschild solution on their exterior, and can possess a photon sphere. 

These are just some examples of possible ambiguities that will generally exist for any non-horizon object whose radial extent makes it fall within the `twilight zone' between the true horizon and the photon sphere. One can always think of a deviation small enough to escape detection. 
In any case, black holes are the most straightforward and general astrophysical interpretation for a range of phenomena. 
For the case of supermassive black holes, the shadow was properly predicted well before the observations with up-to-date computer simulations \citep{falcke2000,broderick2006a,DexterMcKinneyAgol2012a,MoscibrodzkaFalckeShiokawa2016a} and then observed. We have explained here that indeed a black hole's shadow is a clearly defined and robust feature as long as the emission region surrounding the black hole is geometrically thick and optically thin. With these basic assumptions, the exact intensity distribution or details of the astrophysics and particle heating are largely irrelevant. Indeed, for jet sources like M87* it is well established that the core becomes more optically thin at higher frequencies, and that the emission is expected to be spatially extended in all directions. The measured constancy of the shadow diameter of M87* over a decade \citet{WielgusAkiyamaBlackburn2020a} also renders an interpretation of the image as incidental, transient astrophysical circumstances of the accretion disk/jet system unlikely. 

History has shown that it is impossible to derive physics from astrophysics without understanding astrophysics. The crucial question is not whether current or future EHT results depend on astrophysics, but rather how significant the uncertainty due to unknown astrophysics is. The EHT has gone to great lengths to quantify this uncertainty using a wide range of self-consistent numerical simulations that take the behaviour of plasma radiation into account. We have come a long way to understanding how black holes accrete and how they launch jets. Simple toy models, therefore, can only be used to guide our intuition and help us to explain what we are seeing. Nonetheless, care has to be taken that these toy model at least roughly reflect the actual astrophysical situation, which is not entirely arbitrary anymore. 

In summary, we reiterate that the black-hole shadow is a robust and significant observational feature, which speaks of the underlying metric and the darkness of the event horizon, and which is independent of the phenomenon of photon rings. The first detection of a shadow in M87* confirms a basic prediction made for low-power black holes. Of course, this is only a starting point: better tests will require more theoretical investigations, more observations, and better experiments, including space experiments. Uncertainties due to source variability will decrease by repeated EHT observations, and will sharpen constraints on key parameters (see, e.g., \citealt{roelofs2021}, \citealt{PsaltisMedeirosChristian2020a}). Given continuous rubbing of the coin, a more solid outline of the underlying metric will appear, and we will be able to read the mint markings of black holes more clearly in the future.

\section*{Acknowledgments}

TB thanks Jordy Davelaar and Christiaan Brinkerink for insightful comments regarding this research. We thank Prashant Kocherlakota and Hector Olivares for discussion on shadows from black hole impostors, and internal (EHT) referee Maciek Wielgus. This work benefited from discussions in the EHT gravitational physics input working group. 

\bibliography{library.bib}

\end{document}